\documentclass[twocolumn]{aastex631}

\usepackage[utf8]{inputenc}
\usepackage{amsmath}

\newcommand\uJy{$\mu$Jy}
\newcommand\hst{\textit{HST}}
\newcommand\jwst{\textit{JWST}}
\newcommand{\mstar}{$M_\star$}

\newcommand{\mdot}{$M_\odot$}
\newcommand{\lsun}{L$_\odot$}

\newcommand{\sqdeg}{\,deg$^2$}
\newcommand{\sqarcmin}{\,arcmin$^2$}

\newcommand{\cigale}{\texttt{CIGALE}}
\newcommand{\av}{A$_{\rm V}$}
\newcommand{\diver}{\texttt{diver}}

\defcitealias{McKinney2025ApJ...979..229M}{M25}

\shorttitle{SCUBADive II: F150W-Dropouts}
\shortauthors{Manning et al.}

\begin{document}

\title{SCUBADive II: Searching for $z>4$ Dust-Obscured Galaxies via F150W-Dropouts in COSMOS-Web}

\correspondingauthor{Sinclaire M. Manning}
\email{smanning@astro.umass.edu}

\author[0000-0003-0415-0121]{Sinclaire M. Manning}
\affiliation{Department of Astronomy, University of Massachusetts, Amherst, MA 01003, USA}

\author[0000-0002-6149-8178]{Jed McKinney}
\altaffiliation{NASA Hubble Fellow}
\affiliation{Department of Astronomy, The University of Texas at Austin, Austin, TX 78712, USA}

\author[0000-0001-7160-3632]{Katherine E. Whitaker}
\affiliation{Department of Astronomy, University of Massachusetts, Amherst, MA 01003, USA}
\affiliation{Cosmic Dawn Center (DAWN), Denmark}

\author[0000-0002-7530-8857]{Arianna S. Long}
\affiliation{Department of Astronomy, The University of Washington, Seattle, WA 98195, USA}

\author[0000-0003-3881-1397]{Olivia R. Cooper}\altaffiliation{NSF Astronomy and Astrophysics Postdoctoral Fellow}
\affiliation{Department for Astrophysical \& Planetary Science, University of Colorado, Boulder, CO 80309, USA}

\author[0000-0002-0930-6466]{Caitlin M. Casey}
\affiliation{Department of Physics, University of California, Santa Barbara, Santa Barbara, CA 93106, USA}
\affiliation{Cosmic Dawn Center (DAWN), Denmark}

\author[0000-0002-0569-5222]{Rafael C. Arango-Toro}
\affiliation{Aix Marseille Univ, CNRS, CNES, LAM, Marseille, France}

\author[0000-0002-6184-9097]{Jaclyn B. Champagne}
\affiliation{Steward Observatory, University of Arizona, 933 N Cherry Ave, Tucson, AZ 85721, USA}

\author[0000-0003-4761-2197]{Nicole E. Drakos}
\affiliation{Department of Physics and Astronomy, University of Hawaii, Hilo, 200 W Kawili St, Hilo, HI 96720, USA}

\author[0000-0002-9382-9832]{Andreas L. Faisst}
\affiliation{Caltech/IPAC, MS 314-6, 1200 E. California Blvd. Pasadena, CA 91125, USA}

\author[0000-0002-3560-8599]{Maximilien Franco}
\affiliation{Université Paris-Saclay, Université Paris Cité, CEA, CNRS, AIM, 91191 Gif-sur-Yvette, France}

\author[0000-0002-0236-919X]{Ghassem Gozaliasl}
\affiliation{Department of Computer Science, Aalto University, P.O. Box 15400, FI-00076 Espoo, Finland}
\affiliation{Department of Physics, University of, P.O. Box 64, FI-00014 Helsinki, Finland}

\author[0000-0003-0129-2079]{Santosh Harish}
\affiliation{Laboratory for Multiwavelength Astrophysics, School of Physics and Astronomy, Rochester Institute of Technology, 84 Lomb Memorial Drive, Rochester, NY 14623, USA}

\author[0009-0007-3673-4523]{Hossein Hatamnia}
\affiliation{Department of Physics and Astronomy, University of California, Riverside, 900 University Avenue, Riverside, CA 92521, USA}

\author[0000-0003-4073-3236]{Christopher C. Hayward}
\affiliation{Eureka Scientific, Inc., 2452 Delmer Street, Suite 100, Oakland, CA 94602, USA}
\affiliation{Kavli Institute for the Physics and Mathematics of the Universe (WPI), The University of Tokyo Institutes for Advanced Study, The University of Tokyo, Kashiwa, Chiba 277-8583, Japan}

\author[0000-0002-3301-3321]{Michaela Hirschmann}
\affiliation{Institute of Physics, GalSpec, Ecole Polytechnique Federale de Lausanne, Observatoire de Sauverny, Chemin Pegasi 51, 1290 Versoix, Switzerland}
\affiliation{INAF, Astronomical Observatory of Trieste, Via Tiepolo 11, 34131 Trieste, Italy}

\author[0000-0001-9187-3605]{Jeyhan S. Kartaltepe}
\affiliation{Laboratory for Multiwavelength Astrophysics, School of Physics and Astronomy, Rochester Institute of Technology, 84 Lomb Memorial Drive, Rochester, NY 14623, USA}

\author[0000-0002-6610-2048]{Anton M. Koekemoer}
\affiliation{Space Telescope Science Institute, 3700 San Martin Dr., Baltimore, MD 21218, USA}

\author[0000-0001-9773-7479]{Daizhong Liu}
\affiliation{Purple Mountain Observatory, Chinese Academy of Sciences, 10 Yuanhua Road, Nanjing 210023, China}

\author[0000-0002-4872-2294]{Georgios E. Magdis}
\affiliation{Cosmic Dawn Center (DAWN), Jagtvej 128, DK2200 Copenhagen N, Denmark}
\affiliation{DTU-Space, Technical University of Denmark, Elektrovej 327, DK2800 Kgs. Lyngby, Denmark}
\affiliation{Niels Bohr Institute, University of Copenhagen, Jagtvej 128, DK-2200 Copenhagen N, Denmark}

\author[0000-0002-9489-7765]{Henry Joy McCracken}
\affiliation{Institut d’Astrophysique de Paris, UMR 7095, CNRS, and Sorbonne Université, 98 bis boulevard Arago, F-75014 Paris, France}

\author[0000-0002-4485-8549]{Jason Rhodes}
\affiliation{Jet Propulsion Laboratory, California Institute of Technology, 4800 Oak Grove Drive, Pasadena, CA 91001, USA}

\author[0000-0002-4271-0364]{Brant E. Robertson}
\affiliation{Department of Astronomy and Astrophysics, University of California, Santa Cruz, 1156 High Street, Santa Cruz, CA 95064, USA}

\author[0000-0003-4352-2063]{Margherita Talia}
\affiliation{University of Bologna - Department of Physics and Astronomy “Augusto Righi” (DIFA), Via Gobetti 93/2, I-40129 Bologna, Italy}
\affiliation{INAF- Osservatorio di Astrofisica e Scienza dello Spazio, Via Gobetti 93/3, I-40129, Bologna, Italy}

\author[0000-0001-6477-4011]{Francesco Valentino}
\affiliation{Cosmic Dawn Center (DAWN), Denmark}
\affiliation{DTU-Space, Technical University of Denmark, Elektrovej 327, DK2800 Kgs. Lyngby, Denmark}
\affiliation{European Southern Observatory, Karl-Schwarzschild-Str. 2, 85748 Garching, Germany}

\author[0000-0003-1614-196X]{John R. Weaver}\thanks{Brinson Prize Fellow}
\affiliation{MIT Kavli Institute for Astrophysics and Space Research, 77 Massachusetts Ave., Cambridge, MA 02139, USA}

\author[0000-0002-7051-1100]{Jorge A. Zavala}
\affiliation{University of Massachusetts Amherst, 
710 North Pleasant Street, Amherst, MA 01003-9305, USA}


\begin{abstract}
The relative fraction of obscured galaxies at $z>4$ compared to lower redshifts remains highly uncertain as accurate bookkeeping of the dust-obscured component proves difficult. We address this shortcoming with SCUBADive, a compilation of the \jwst\ counterparts of (sub-)millimeter galaxies in COSMOS-Web, in order to further analyze the distribution and properties of massive dust-obscured galaxies at early times. In this paper, we present a subset of SCUBADive, focusing on 60 ``dark'' galaxies that dropout at 1.5\micron. Motivated by \jwst\ observations of AzTECC71, a far-infrared bright F150W-dropout with $z_{\rm phot}=5.7^{+0.8}_{-0.7}$, we complete a systematic search of F150W-dropouts with SCUBA-2 and ALMA detections to find more candidate high redshift dusty galaxies. Within our subsample, 16 are most similar to AzTECC71 due to fainter F444W magnitudes ($>24$\,mag) and lack of counterparts in COSMOS2020. Despite high star formation rates ($\langle$SFR$\rangle=450^{+920}_{-320}$\,\mdot\,yr$^{-1}$) and large stellar masses ($\langle$log$_{10}$(\mstar)$\rangle=11.2^{+0.5}_{-0.6}$\,\mdot) on average, these galaxies may not be particularly extreme for their presumed epochs according to offsets from the main sequence. We find that heavily obscured galaxies, which would be missed by pre-\jwst\ optical imaging campaigns, comprise $\gtrsim20$\% of galaxies across mass bins and potentially contribute up to 60\% at the very high mass end (log$_{10}$(\mstar/\mdot)$>11.5$) of the $z>4$ stellar mass function. 
\end{abstract}

\keywords{galaxies: evolution -- galaxies: high-redshift -- submillimeter: galaxies -- methods: observational}

\section{Introduction}
Galaxies that shine brightly at far-infrared (FIR) and (sub-)millimeter (mm) wavelengths (L$_{\rm IR}>10^{11}$\,\lsun) do so thanks to their large reservoirs of cosmic dust \citep*[M$_{\rm dust}>10^8$\mdot;][]{Casey2014PhR...541...45C}; silicate and carbonaceous grains which absorb the ultraviolet (UV) and optical light from young stars and re-radiate it at longer wavelengths. This fundamental characteristic, which allows for easy identification in the FIR/(sub-\!)mm regime, is the source of uncertainty for many fundamental properties of these galaxies as the dearth of rest-frame UV/optical detections leads to large inaccuracies and discrepancies in spectral energy distribution (SED) modeling. Heavily dust-obscured galaxies are certainly not a new class of object, having been detected by single-dish facilities operating at FIR/mm wavelengths for the past several decades \citep{Smail1997ApJ...490L...5S,Barger1998Natur.394..248B,Hughes1998Natur.394..241H,Eales1999ApJ...515..518E}, but the ability to observe and fully characterize them with higher resolution and sensitivity was recently enabled with newer facilities like the Atacama Large Millimeter/submillimeter Array (ALMA) and \jwst.

Among the well-studied population of dusty galaxies exists a sub-population of the most obscured sources (A$\mathrm{_V} \gtrsim 2$) characterized by high redshift solutions ($z>3$) and non-detections in available ground and space-based rest-frame UV/optical observations. They have thus been largely missed by surveys with selection methods reliant on detections at shorter wavelengths. Again, although such extreme dusty sources have been identified in the past \citep{Dannerbauer2002ApJ...573..473D,Chapman2002ApJ...570..557C,Frayer2004AJ....127..728F,Smail2004ApJ...616...71S,Pope2005MNRAS.358..149P,Simpson2014ApJ...788..125S}, the last several years have seen a surge in new identifications of these ``dark'' galaxies.  Their presence is made more intriguing given the impressive depths reached by more recent UV/optical observations, for example with the \textit{Hubble Space Telescope} (\hst), as compared to previous studies. 

Various names exist for this sub-population, stemming from the range of instruments, wavelength coverage, and survey depths available to individual studies (\hst-dark; optical/near-infrared (OIR)-dark; ALMA-only; H/K-band dropouts; optically faint galaxies (OFGs); \citealt{Wang2016ApJ...816...84W,Franco2018A&A...620A.152F,Schreiber2018aA&A...611A..22S,Wang2019Nature,Williams2019ApJ...884..154W,Yamaguchi2019ApJ...878...73Y,Talia2021ApJ...909...23T,Smail2021MNRAS.502.3426S,Manning2022ApJ...925...23M,Perez-Gonzalez2023ApJ...946L..16P,Barrufet2023MNRAS.522..449B,Gomez-Guijarro2023A&A...677A..34G}). Interestingly, these works have shown the population of ``dark'' galaxies to encompass a broad range of stellar masses ($M_\star$) and star-formation rates (SFRs), spanning extreme starbursts \citep{Walter2012Natur.486..233W} with flux densities akin to canonical submillimeter galaxies and dusty star-forming galaxies (SMGs/DSFGs; \citealt{Blain2002PhR...369..111B,Ivison2010MNRAS.402..245I,Casey2014PhR...541...45C}), to more normal-type star-forming galaxies \citep{LeFevre2020A&A...643A...1L,Fudamoto2021Natur.597..489F}. Additionally, a handful have been spectroscopically confirmed to exist within the first billion and a half years after the Big Bang \citep[e.g. $z>4-6$;][]{Strandet2017ApJ...842L..15S,Marrone2018Natur.553...51M,Jin2019ApJ...887..144J,Casey2019ApJ88755C,Xiao2024Natur.635..311X,Herard-Demanche2025MNRAS.537..788H,Barrufet2025MNRAS.tmp...22B}. 

The Venn diagram of ``dark'' galaxies, i.e. how samples overlap and which distinct galaxy populations exist within them, has become increasingly hard to disentangle. Simply determining a common threshold to define darkness, e.g. a single magnitude cutoff, is not so straightforward. Samples are being collected incrementally as data becomes available and direct comparison between samples proves challenging due to varying survey depths and wavelength coverage. Having said that, extricating the makeup of ``dark'' galaxies is garnering more interest due to their suggested prevalence. Depending on the employed selection techniques and surveys, anywhere from $20-50$\% of dusty galaxies are ``dark'' \citep{DaCunha2015ApJ806110D,Simpson2017ApJ...839...58S,Dudzevi2020MNRAS.494.3828D, Manning2022ApJ...925...23M}, and thus potentially omitted from galaxy evolution studies based on optical imaging alone. Given that dust-obscured galaxies tend to be the most massive galaxies at their epoch \citep{DaCunha2015ApJ806110D,Miettinen2017A&A...606A..17M} and that stellar mass is known to correlate with obscuration \citep{Whitaker2017ApJ...850..208W}, including this population will impact the shape of the evolving stellar mass function \citep{Deshmukh2018ApJ...864..166D,Long2023ApJ...953...11L}, dark matter halo clustering strengths \citep{Lim2020ApJ...895..104L,GarciaVergara2020ApJ...904....2G}, and offer insights into star formation efficiency in the early Universe \citep{Stach2021MNRAS.504..172S,Rybak2022A&A...667A..70R}. The advent of large ALMA surveys with comparable resolutions to \hst/\jwst\ tracing dust emission in conjunction with those from \jwst\ probing previously hidden stellar components, means we can begin to make significant strides in the search for these systems, their subsequent characterization, and ultimately their impact on galaxy evolution models. 

The work of \cite{McKinney2023ApJ...956...72M} reported the serendipitous discovery of the first F150W-dropout dusty galaxy among the initial \jwst\ NIRCam imaging (6/152 visits, 77\sqarcmin) acquired from the COSMOS-Web survey \citep{Casey_Kartaltepe2023ApJ...954...31C}. The source, identified in the literature as AzTECC71 \citep{Aretxaga2011,Brisbin2017A&A...608A..15B,Simpson2019ApJ88043S}, has robust cold dust continuum detections ($\lambda\geq850\,{\micron}$) from several (sub-)millimeter facilities, but no detections at observed wavelengths shorter than 4\,\micron. AzTECC71 is believed to be undergoing an extreme burst of star formation (SFR$\approx800\,M_\odot$\,yr$^{-1}$) with considerable stellar and dust mass already built up ($M_\star\approx4\times10^{10}\,M_\odot$ and $M_{\rm dust}\approx10^{8}\,M_\odot$, respectively), as is typical for SMG/DSFGs, but with an intriguingly high photometric redshift solution ($z_{\rm phot}\approx5.7$). The re-discovery of AzTECC71 presented a potential path forward to identifying $z>4$ DSFGs in order to address questions regarding their prevalence in the first two billion years of the Universe \citep{Casey2018aApJ...862...77C}. The remaining visits of the COSMOS-Web survey have since been completed, so we follow-up the initial finding of AzTECC71 with a systematic search for more distant, dust-obscured, ``dark'' galaxies in the full 0.54\sqdeg\ COSMOS-Web survey area.


In this paper, we present 60 heavily dust-obscured galaxies that, like AzTECC71, dropout in the F150W filter of COSMOS-Web and are detected by SCUBA-2/JCMT with high photometric redshift solutions. Section \ref{sec:data} describes the data products used for this analysis, including a description of the broader SCUBADive project and the selection of the F150W-dropout sample. Section \ref{sec:seds and photzs} reports the results from spectral energy distribution fitting. Section \ref{sec:results} discusses derived physical properties and the utility of employing an F150W-dropout selection to identify distant, dusty galaxies like AzTECC71. In Section \ref{sec:discussion}, we present a comparison with other selection methods for high redshift red galaxies and discuss the nature of the sources identified. Section \ref{sec:conclusions} summarizes our findings. Throughout this work we assume a \cite{Chabrier2003PASP..115..763C} initial mass function (IMF), AB magnitudes, and a $\Lambda$CDM cosmology with $H_0=70$\,km\,s$^{-1}$\,Mpc$^{-1}$, $\Omega_m=0.3$, and $\Omega_\Lambda=0.7$. 

\section{Data}\label{sec:data}
This work presents an analysis of dust-obscured galaxies found in the \jwst\-observed COSMOS-Web area of the Cosmic Evolution Survey Field (COSMOS; \citealt{Capak2007ApJS..172...99C,Scoville2007ApJS..172....1S}); one of the richest extragalactic legacy datasets available with deep and wide coverage from X-ray to radio wavelengths and over 30 bands of optical/near-IR imaging. Here, we summarize the data most relevant for the creation of the SCUBADive sample of 289 SMG/DSFGs (described in detail in \cite{McKinney2025ApJ...979..229M}, hereafter \citetalias{McKinney2025ApJ...979..229M}) and our follow-up investigation of 60 \jwst\ F150W-dropouts. With the SCUBADive sample (\citetalias{McKinney2025ApJ...979..229M}), we aim to characterize every SCUBA-2 SMG/DSFG in COSMOS-Web that has ALMA and \jwst\ imaging.

The construction of the flux-limited SCUBADive parent catalog began with 1985 known submillimeter sources ($S_{850\micron}>2$\,mJy) observed with the SCUBA-2 instrument on the James Clerk Maxwell Telescope (JCMT) as part of the Cosmology Legacy Survey (S2CLS; \citealt{Geach2017}) and follow-up SCUBA-2 COSMOS survey (S2COSMOS; \citealt{Simpson2019ApJ88043S}). From these surveys, 706 SCUBA-2 sources, representing $\sim36$\% of the parent catalog, were found to overlap with the COSMOS-Web NIRCam area.

In the case of single dish, low resolution SCUBA-2 observations, attempting to associate the correct optical/near-IR counterpart to an unresolved submillimeter source is known to be challenging due to the proclivity of the large beam size to span across multiple galaxies \citep{Pope2008ApJ...675.1171P}, so the retrieval of additional high resolution archival data from ALMA was critical for counterpart identification with \jwst\ imaging from COSMOS-Web. All ALMA observations in Band 6 and/or 7 between $800-1250$\,\micron\ (falling within 15\arcsec\ of a SCUBA-2 source and inside the COSMOS-Web NIRCam area) were downloaded from the ALMA archive. \footnote{The collated ALMA dataset reflects all observations publicly available on the archive prior to September 7th, 2023.} All data were then uniformly re-imaged with the Common Astronomy Software Application (\texttt{CASA 6.6.3}; \citealt{McMullin2007ASPC..376..127M}) and its deconvolution task, \texttt{tclean}, using natural weighting. Source extraction was carried out using \texttt{PyBDSF} \citep{MohanRafferty2015ascl.soft02007M} and, after the removal of spurious sources, \citetalias{McKinney2025ApJ...979..229M} established a final SCUBADive sample of 289 galaxies (corresponding to 219 SCUBA-2/850\micron\ sources). 

With high resolution ALMA data in hand, sources are then cross-matched to the 0.54\sqdeg\ \jwst\ Cosmic Origins Survey (COSMOS-Web; PID \#1727, PIs Casey \& Kartaltepe, \citealt{Casey_Kartaltepe2023ApJ...954...31C}).  
We make use of the substantial ancillary ground- and space-based data available through the COSMOS Survey to aid in spectral energy distribution modeling. Radio and/or X-ray observations are given special consideration as detected emission at these wavelengths may suggest the presence of active galactic nuclei (AGN).
SCUBADive folds in radio data from the Karl G. Jansky Very Large Array (VLA) at 1.4\,GHz and 3\,GHz as part of the VLA-COSMOS Survey ($3\sigma\sim30\,\mu$Jy\,beam$^{-1}$; \citealt{Schinnerer2010ApJS..188..384S}) and VLA-COSMOS 3\,GHz Large Project ($5\sigma\sim12\,\mu$Jy\,beam$^{-1}$; \citealt{Smolcic2017}). 
Of the F150W-dropout sample, 40\% (24/60) do not have radio counterparts, at least to the depths of the current observations in COSMOS, and 75\% (12/16) of the faintest sources (F444W\,$>24$\,mag) lack detections, including AzTECC71. 

\subsection{Sample Selection: Identifying F150W-Dropouts from SCUBADive}\label{sec:selection}
Following the successful \jwst\ counterpart identification of AzTECC71 \citep{McKinney2023ApJ...956...72M} in the initial 77.76\,\sqarcmin\ COSMOS-Web data release from January 2023 \citep{Franco2023arXiv230800751F} and the subsequent creation of the SCUBADive \jwst$+$ALMA catalog across all $0.54$\sqdeg\ of the full COSMOS-Web Survey (\citetalias{McKinney2025ApJ...979..229M}), we now search for more F150W-dropout dust-obscured galaxies. From the SCUBADive sample of 289 galaxies we down-select to only include those which satisfy the following criteria: non-detections ($<4\,\sigma$) in the F115W and F150W \jwst\ NIRCam filters -- with $5\,\sigma$ depths of 27.45 and 27.66 AB magnitudes, respectively -- and ($>4\,\sigma$) detections in F444W. These criteria were chosen to identify galaxies with similar properties to AzTECC71, as our search is focused on finding the highest redshift SMGs/DSFGs. In total, 60 galaxies are classified as F150W-dropouts and these sources are the focus of this work. 

\section{Photometry and SED Fitting}\label{sec:seds and photzs}

\subsection{Aperture Photometry with \texttt{diver}}
\begin{figure*}
    \centering
    \includegraphics[width=\textwidth,trim={0 0 0 5.5mm}]{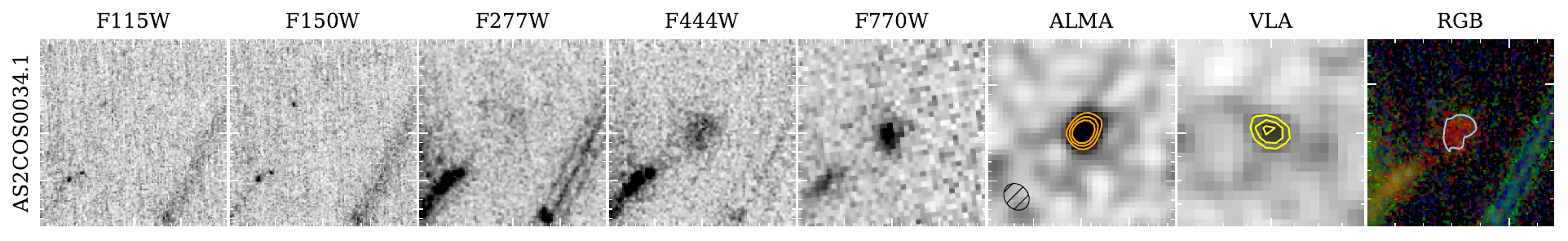}
    \includegraphics[width=\textwidth,trim={0 0 0 5.5mm}]{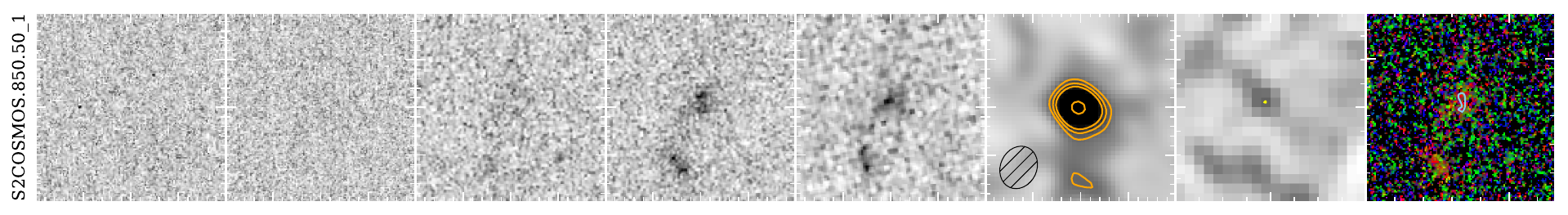}
    \includegraphics[width=\textwidth,trim={0 0 0 5.5mm}]{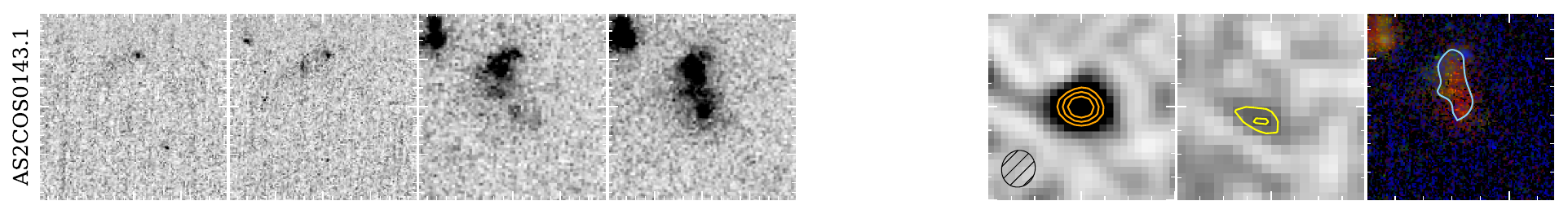}
    \includegraphics[width=\textwidth,trim={0 0 0 5.5mm}]{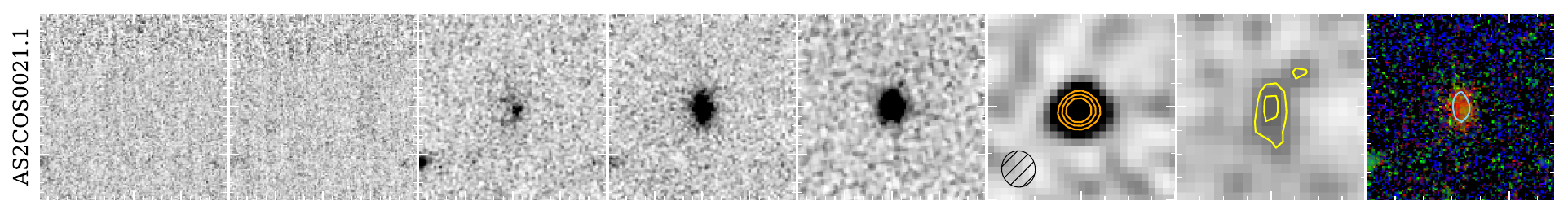}
    \includegraphics[width=\textwidth,trim={0 0 0 5.5mm}]{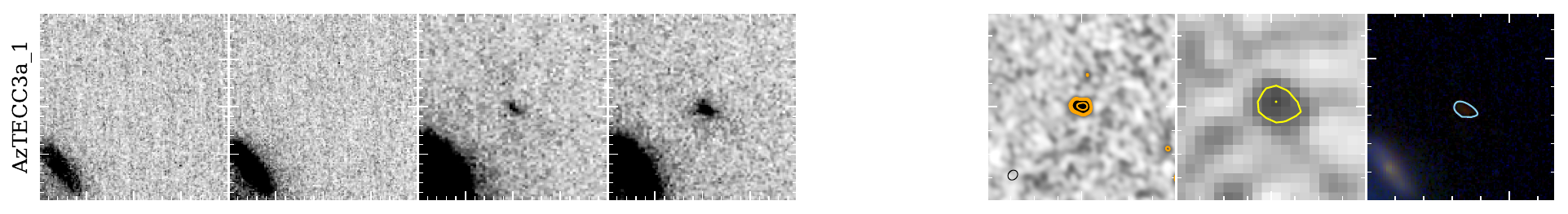}
    \includegraphics[width=\textwidth,trim={0 0 0 5.5mm}]{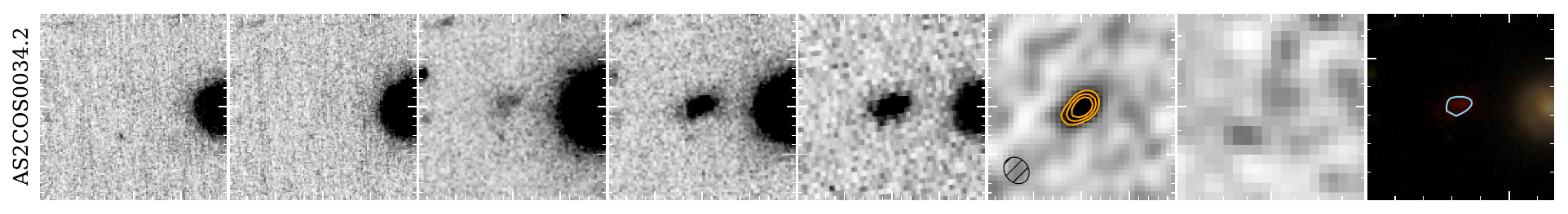}
    \includegraphics[width=\textwidth,trim={0 0 0 5.5mm}]{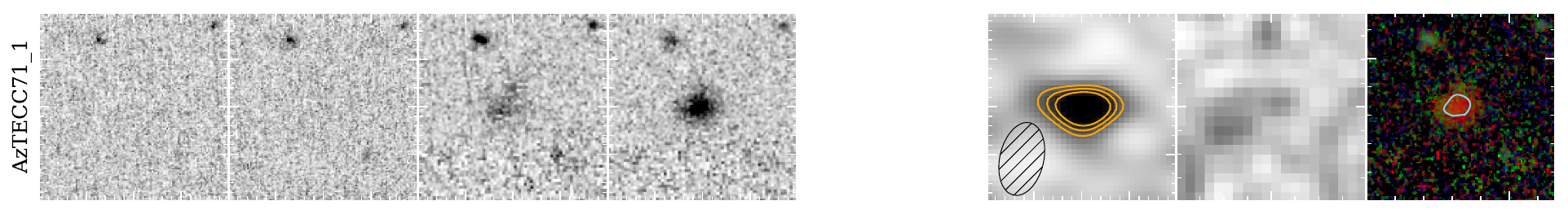}
    \includegraphics[width=\textwidth,trim={0 0 0 5.5mm}]{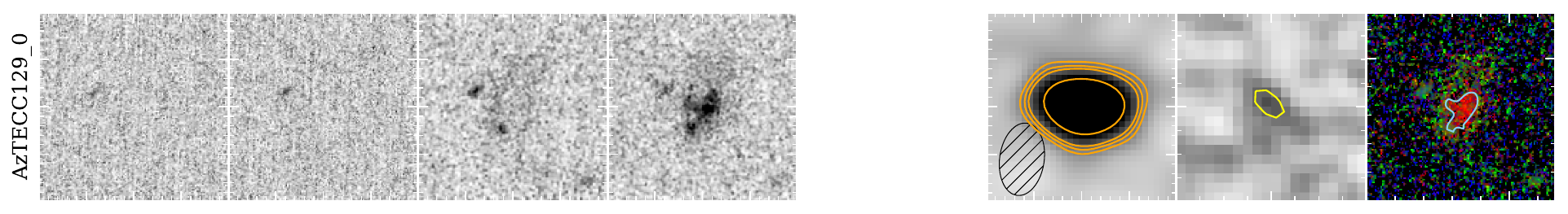}
    \caption{4\arcsec$\times$4\arcsec\ cutouts of the 16 faintest (in F444W) F150W-dropouts found via SCUBA-Diving in COSMOS-Web. Sources lacking F770W imaging fall outside of the MIRI coverage in the field. Orange and yellow contours represent SNR of 3, 4, 5, 10$\sigma$ in their respective ALMA and VLA cutouts. The hatched ellipse represents the beam size of the given ALMA data. The rightmost column shows a combined RGB image of F115W$+$F150W/F277W/F444W with the custom \diver\ aperture shown in light blue. For a comprehensive view and discussion of the known merger S2COSMOS.850.50\_1 (a.k.a. MAMBO-9), we refer the reader to \cite{Casey2019ApJ88755C} and \cite{Akins2025arXiv250806607A}.} 
    \label{fig:cutouts}
\end{figure*}

Photometry is derived for the SCUBADive sample via the custom aperture photometry code \diver. A full description on the methodology of \diver\ can be found in Section 5.1 of \citetalias{McKinney2025ApJ...979..229M} and is briefly summarized here. The novelty of \diver\ lies in its use of the \jwst\ resolution as it takes the reddest NIRCam image containing the source (most often F444W) and creates a custom, non-parametric aperture to measure flux densities in all available imaging from COSMOS on native-resolution. This ensures the full extent of the source is encompassed in the aperture with limited additional noise. For imaging with poorer spatial resolution, PSF correction factors determined from the ratio of the total convolved flux to the convolved flux interior to the custom aperture are applied. Flux errors are determined from a combination of sky noise and detector level uncertainties (see Equation 1 in \citetalias{McKinney2025ApJ...979..229M}). Examples of these custom apertures can be seen in Figure \ref{fig:cutouts}. 

\begin{figure*}
    \centering
    \includegraphics[width=\textwidth,trim={0 0 0 5.5mm}]{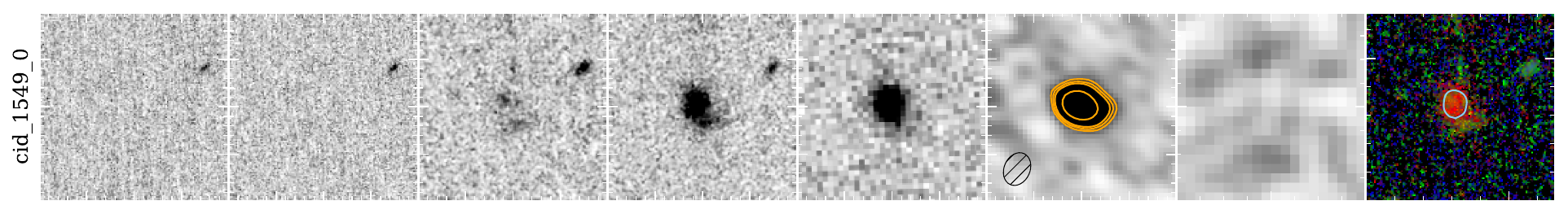}
    \includegraphics[width=\textwidth,trim={0 0 0 5.5mm}]{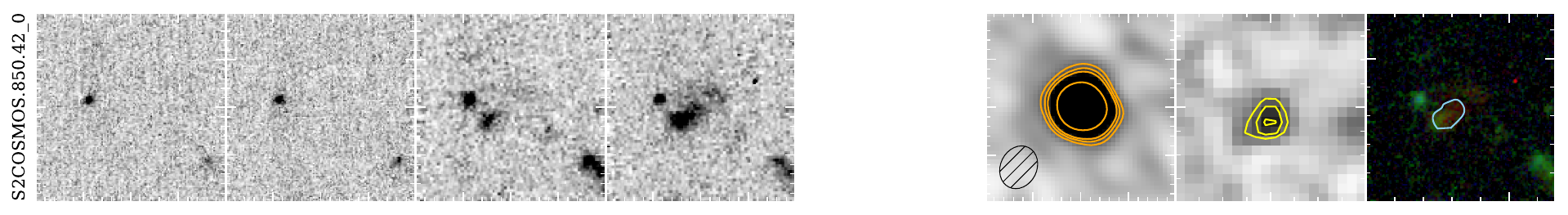}
    \includegraphics[width=\textwidth,trim={0 0 0 5.5mm}]{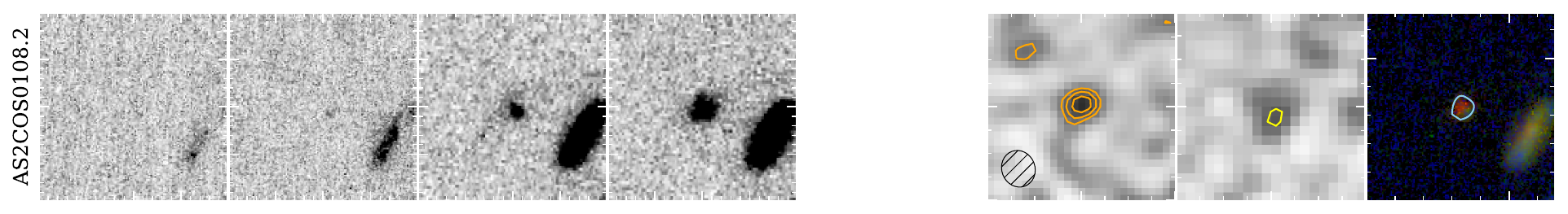}
    \includegraphics[width=\textwidth,trim={0 0 0 5.5mm}]{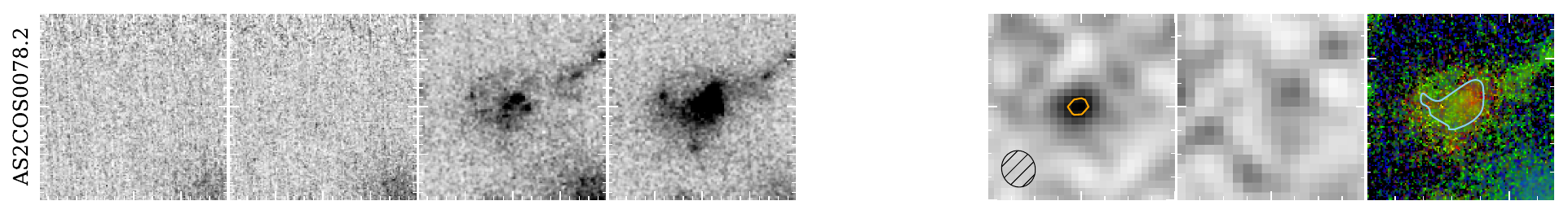}
    \includegraphics[width=\textwidth,trim={0 0 0 5.5mm}]{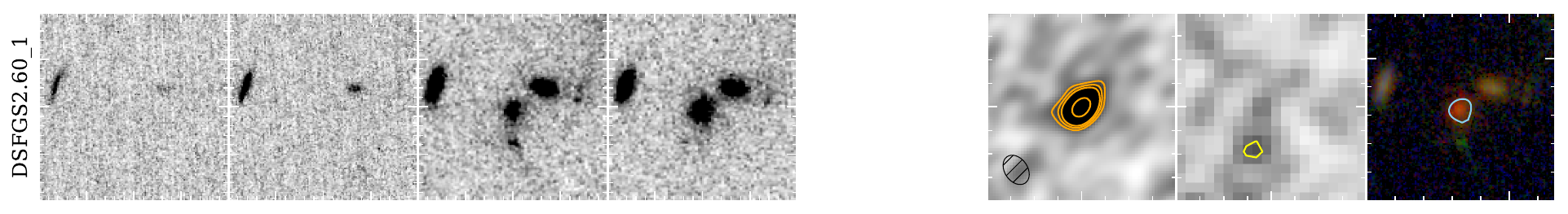}
    \includegraphics[width=\textwidth,trim={0 0 0 5.5mm}]{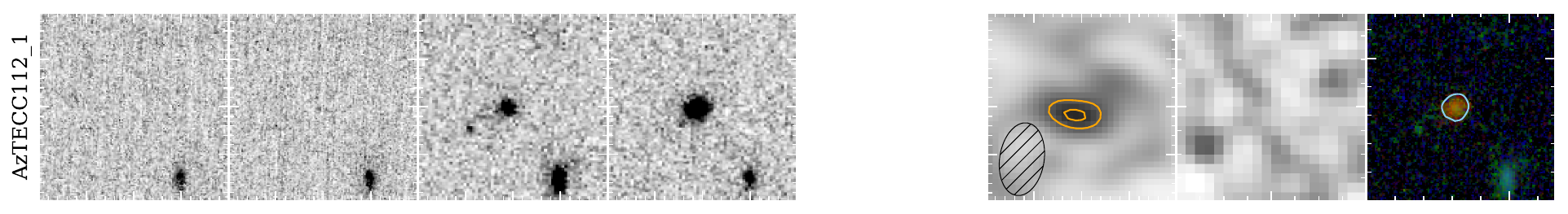}
    \includegraphics[width=\textwidth,trim={0 0 0 5.5mm}]{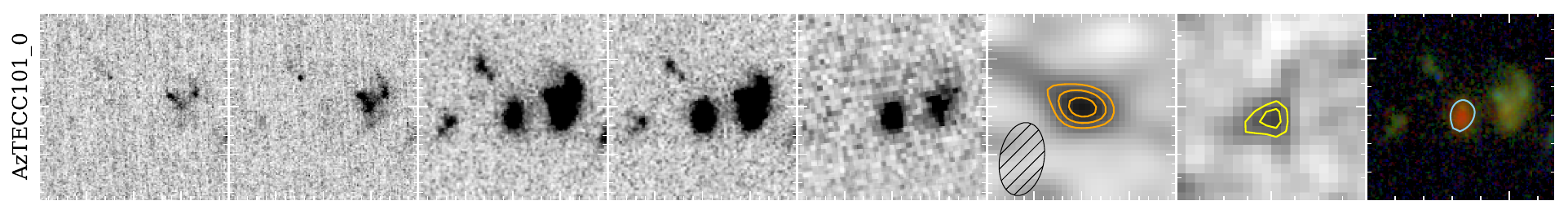}
    \includegraphics[width=\textwidth,trim={0 0 0 5.5mm}]{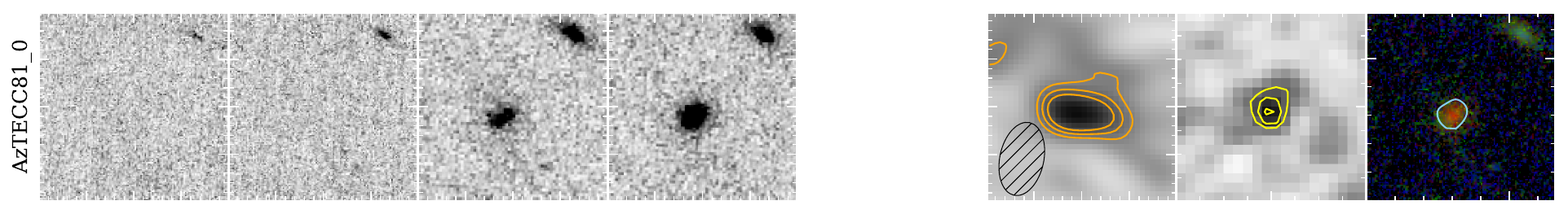}
\end{figure*}

\begin{deluxetable*}{lCCCCCCC}
\tablecaption{\jwst\ and IRAC Photometry \label{tab:jwst_data}}
\tablehead{
\colhead{Name}&
\colhead{F115W}&
\colhead{F150W}&
\colhead{F277W}&
\colhead{F444W}&
\colhead{F770W}&
\colhead{IRAC Ch1}&
\colhead{IRAC Ch2} \\
\colhead{} & \colhead{[nJy]} & \colhead{[nJy]} & \colhead{[nJy]} & \colhead{[nJy]} & \colhead{[\uJy]} & \colhead{[\uJy]} & \colhead{[\uJy]}
}
\startdata
AS2COS0034.1 & (9.9\pm3.3) & (8.2\pm2.7) & 20.8\pm1.1 & 23.9\pm1.6 & 0.090\pm0.002 & 10.1\pm0.7 & 6.91\pm0.49 \\
S2COSMOS.850.50\_1 & (3.7\pm7.6) & (4.3\pm6.1) & (5.4\pm3.0) & 36.3\pm3.9 & -- & (0.11\pm0.16) & (0.36\pm0.12) \\
AS2COS0143.1 & (2.0\pm8.1) & (0.4\pm6.2) & 26.1\pm2.7 & 106.2\pm3.8 & -- & 3.03\pm0.22 & 3.70\pm0.23 \\
AS2COS0021.1 & (0.2\pm8.1) & (-3.8\pm6.5) & 30.9\pm3.1 & 125.8\pm4.1 & 0.79\pm0.02 & (0.45\pm0.21) & (0.64\pm0.17) \\
AzTECC3a\_1 & (-8.3\pm15.3) & (7.5\pm13.0) & 79.5\pm6.6 & 193.7\pm8.8 & -- & 1.91\pm0.44 & 1.73\pm0.26 \\
AS2COS0034.2 & (9.0\pm11.4) & (14.2\pm9.5) & 48.0\pm5.8 & 210.8\pm5.5 & 0.81\pm0.02 & 4.88\pm0.50 & 4.03\pm0.32 \\
AzTECC71\_1 & (12.5\pm18.2) & (11.5\pm16.9) & 63.8\pm9.2 & 227.5\pm10.4 & -- & (0.34\pm0.26) & (0.57\pm0.28) \\
AzTECC129\_0 & (0.0\pm23.6) & (-3.2\pm21.2) & 50.9\pm10.6 & 282.0\pm14.2 & 1.73\pm0.04 & (0.49\pm0.19) & (0.76\pm0.20) \\
cid\_1549\_0 & (-6.2\pm19.4) & (-17.6\pm15.8) & 56.1\pm10.1 & 373.1\pm10.2 & 1.83\pm0.04 & (0.26\pm0.30) & (0.80\pm0.27) \\
S2COSMOS.850.42\_0 & (24.1\pm28.6) & (6.0\pm22.3) & 138.7\pm11.6 & 423.1\pm13.3 & -- & 0.78\pm0.19 & 0.86\pm0.17 \\
AS2COS0108.2 & (19.5\pm11.4) & (17.3\pm9.4) & 137.7\pm4.3 & 547.9\pm8.0 & -- & 1.38\pm0.22 & 2.03\pm0.16 \\
AS2COS0078.2 & (-16.1\pm17.8) & (3.9\pm14.0) & 195.9\pm8.2 & 620.1\pm11.8 & -- & 3.78\pm0.35 & 3.76\pm0.24 \\
DSFGS2.60\_1 & (7.3\pm23.5) & (14.1\pm19.3) & 260.7\pm8.5 & 642.4\pm9.9 & -- & 2.03\pm0.33 & 1.63\pm0.24 \\
AzTECC112\_1 & (-36.3\pm34.0) & (-8.9\pm26.0) & 228.3\pm9.9 & 726.5\pm13.8 & -- & (0.66\pm0.19) & 1.06\pm0.18 \\
AzTECC101\_0 & (-16.8\pm23.1) & (54.2\pm18.3) & 301.7\pm8.6 & 775.0\pm9.2 & 1.87\pm0.03 & 1.31\pm0.16 & 1.85\pm0.15 \\
AzTECC81\_0 & (42.4\pm31.2) & (50.3\pm23.2) & 280.5\pm16.7 & 810.8\pm16.5 & -- & (0.50\pm0.35) & 1.15\pm0.25 \\
\enddata
\tablecomments{\jwst\ and IRAC photometry of the 16 faintest (F444W$>24$\,mag) F150W-dropout galaxies in SCUBADive reverse ordered by their F444W magnitudes. Non-detections and measurements with $<3.5\sigma$ significance are denoted by parenthesis.} 
\end{deluxetable*}

\begin{deluxetable*}{llCCCC}
\tablecaption{Millimeter and Radio Flux Densities \label{tab:fir_data}}
\tablehead{
\colhead{Name}&
\colhead{Position}&
\colhead{$S_{\rm 850}$$^{(a)}$}& 
\colhead{$S_{\rm 870}$}& 
\colhead{$S_{\rm 1.3\,mm}$}& 
\colhead{$S_{\rm 3\,GHz}$$^{(b)}$} \\
\colhead{} & \colhead{} & \colhead{[mJy]} & \colhead{[mJy]} & \colhead{[mJy]} & \colhead{[$\mu$Jy]}
}
\startdata
AS2COS0034.1 & 10:00:25.070 +02:26:07.32 & 6.2\pm0.1 & 5.29\pm0.02 & -- & 13.0\pm2.4 \\
S2COSMOS.850.50\_1 & 10:00:26.352 +02:15:27.88 & 5.9\pm0.6 & -- & 1.45\pm0.14 & -- \\
AS2COS0143.1 & 09:59:49.940 +01:55:13.79 & 7.5\pm0.1 & 5.50\pm0.02 & -- & -- \\
AS2COS0021.1 & 10:00:24.151 +02:20:05.37 & 6.6\pm0.2 & 6.59\pm0.01 & -- & -- \\
AzTECC3a\_1 & 10:00:07.847 +02:26:13.28 & 16.8\pm0.6 & 1.94\pm0.15 & -- & -- \\
AS2COS0034.2 & 10:00:25.371 +02:26:05.32 & 6.2\pm0.1 & 3.75\pm0.01 & -- & -- \\
AzTECC71\_1 & 09:59:52.954 +02:18:49.13 & 3.6\pm0.7 & -- & 2.13\pm0.26 & -- \\
AzTECC129\_0 & 10:01:30.168 +02:02:13.90 & 5.2\pm1.0 & -- & 2.69\pm0.12 & 8.2\pm2.3 \\
cid\_1549\_0 & 09:59:44.022 +02:21:08.83 & (3.4\pm1.0) & 2.45\pm0.14 & -- & -- \\
S2COSMOS.850.42\_0 & 10:00:06.485 +02:38:37.64 & 9.1\pm1.0 & -- & 3.74\pm0.17 & 12.6\pm2.4 \\
AS2COS0108.2 & 10:00:22.811 +01:51:36.68 & 6.9\pm0.1 & 1.01\pm0.01 & -- & -- \\
AS2COS0078.2 & 10:01:47.483 +02:24:55.11 & 9.5\pm0.1 & 3.36\pm0.01 & -- & -- \\
DSFGS2.60\_1 & 10:00:00.622 +02:27:35.74 & (2.0\pm1.1)& 1.61\pm0.13 & -- & -- \\
AzTECC112\_1 & 10:00:10.788 +01:53:05.30 & (1.9\pm1.1) & -- & 0.47\pm0.12 & (-0.1\pm4.8) \\
AzTECC101\_0 & 09:59:45.871 +02:30:25.15 & (2.1\pm1.1) & -- & 0.87\pm0.15 & 11.1\pm2.3 \\
AzTECC81\_0 & 10:00:06.276 +01:52:48.14 & 5.1\pm1.0 & -- & 1.0\pm0.2 & 12.5\pm2.4 \\
\enddata
\tablecomments{Positions and flux densities from available (sub-)millimeter and radio data of the faintest F150W-dropout galaxies in SCUBADive. $^{(a)}$850\micron\ fluxes are from \cite{Simpson2019ApJ88043S}. In cases where the 850 fluxes are less than $4\sigma$, this means the original robust detection has been deblended between multiple ALMA sources within the beam, sometimes resulting in lower SNRs. $^{(b)}$Radio fluxes from the VLA-COSMOS Survey \citep{Smolcic2017}. The positions reported correspond to the peak pixel of their associated ALMA detections. Values in parenthesis indicate $<3.5\sigma$ detections.}
\end{deluxetable*}

\subsection{SED Fitting}
\label{subsec:SED_Fitting}
Spectral energy distributions (SEDs) for the 60 F150W-dropout galaxies were modeled using the \texttt{python} version of the Code Investigating GALaxy Emission, (\cigale; \citealt{Boquien2019A&A...622A.103B}) as part of the larger SCUBADive survey in \citetalias{McKinney2025ApJ...979..229M}. \cigale\ was chosen over the myriad of other SED fitting codes for its inclusion of FIR and radio data rather than only fitting the UV to NIR regime. This is key for accurately modeling the physical properties of dust-obscured galaxies. A full description of the SED modeling along with a comparison to results from
\texttt{magphys} \citep{DaCunha2008MNRAS3881595D,DaCunha2015ApJ806110D,Battisti2019ApJ...882...61B} can be found in Section 6.2 of \citetalias{McKinney2025ApJ...979..229M}. We list the selected parameters and adopted values used in \texttt{CIGALE} in Table \ref{tab:cigale} for clarity and reproducibility. Included in this table are the parameters for the inclusion of an AGN component not originally explored in \citetalias{McKinney2025ApJ...979..229M} as we opt to run \cigale\ a second time to test for potential AGN contamination in the F150W-dropout sample. As concluded in \citetalias{McKinney2025ApJ...979..229M} for the entire SCUBADive sample, we also find that AGN do not significantly contaminate the F150W-dropouts. Three sources (AzTECC159\_0, AzTECC91\_1, and AS2COS0198.1) were found to have potentially non-negligible fractional AGN contributions to their total IR luminosity ($f_{\rm AGN}>0.1$), but none exhibit point-like morphologies. Given these findings and lack of appropriate data to disentangle AGN from star formation, all derived properties come from the original \cigale\ run without the AGN component included. Mid-IR observations to detect emission from the hot torus dust and/or deeper X-ray data will be needed to place further constraints on the presence of AGN in these galaxies.

As explained in \citetalias{McKinney2025ApJ...979..229M}, photometric redshifts for SCUBADive galaxies are derived from fitting just the optical-NIR photometry up to $\sim8$\,\micron\ as the results were shown to better recover known spectroscopic redshifts. These photometric redshifts ($z_{\rm opt}$) are then adopted as the fiducial redshift for a full multiwavelength fit to determine physical properties. However, it is also noted in \citetalias{McKinney2025ApJ...979..229M} that catastrophic failures for SMGs/DFGs can occur when omitting the FIR/radio data (also demonstrated in \citealt{Battisti2019ApJ...882...61B}). In the case of the F150W-dropout subsample, two sources (AS2COS0034.1 and S2COSMOS850.42.0) were misidentified as $z\sim0.5$ dwarf galaxies (\mstar$\sim10^{6}$\mdot) in \citetalias{McKinney2025ApJ...979..229M}. For AS2COS0034.1, we instead report the photometric redshift and physical properties derived from the full panchromatic SED fit. In the case of S2COSMOS850.42.0, a spectroscopically confirmed redshift prevents this specific modeling failure.

\begin{figure*}
    \centering
    \includegraphics[scale=0.33]{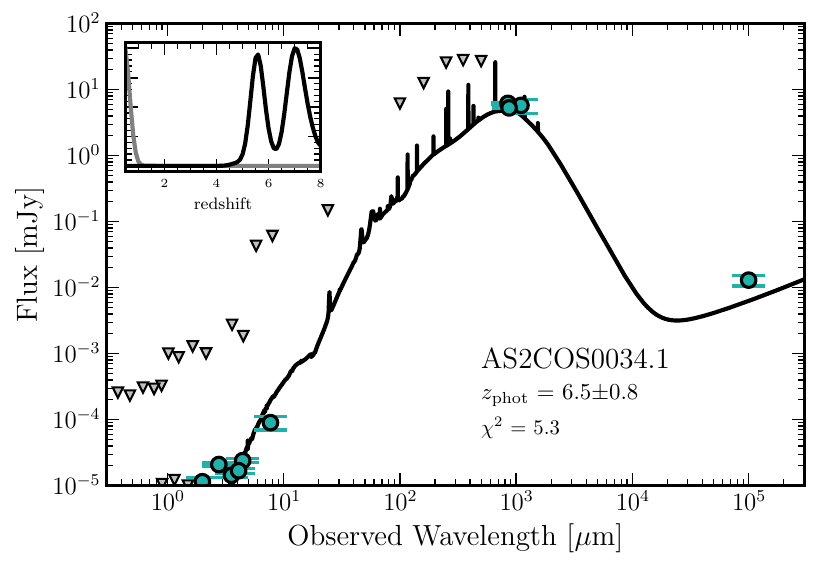}\includegraphics[scale=0.33]{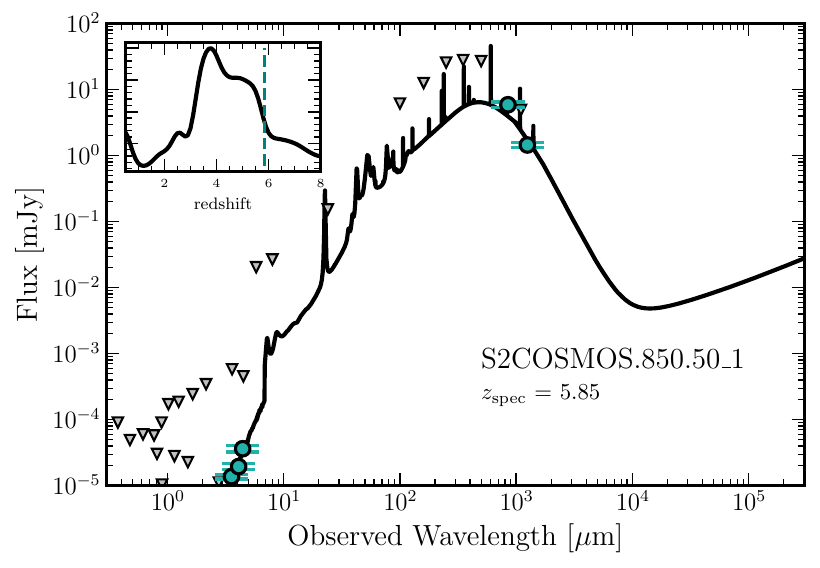}\includegraphics[scale=0.33]{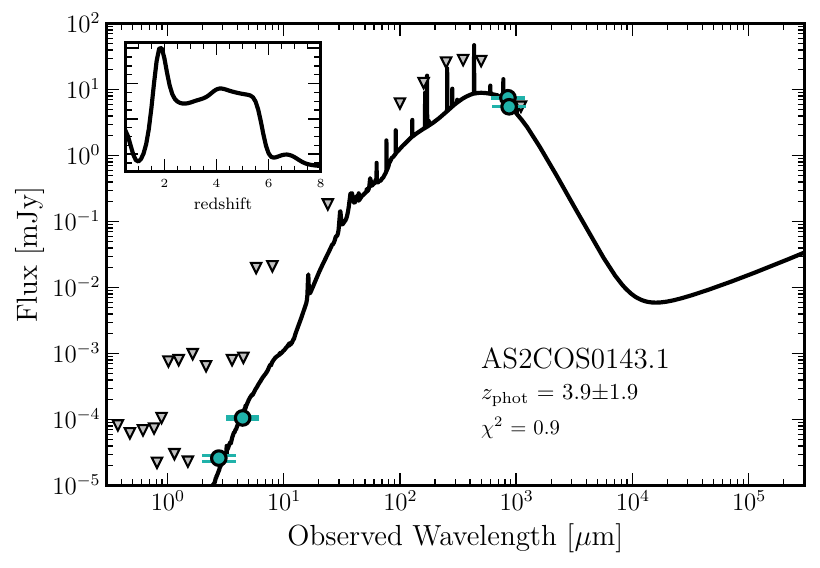}\includegraphics[scale=0.33]{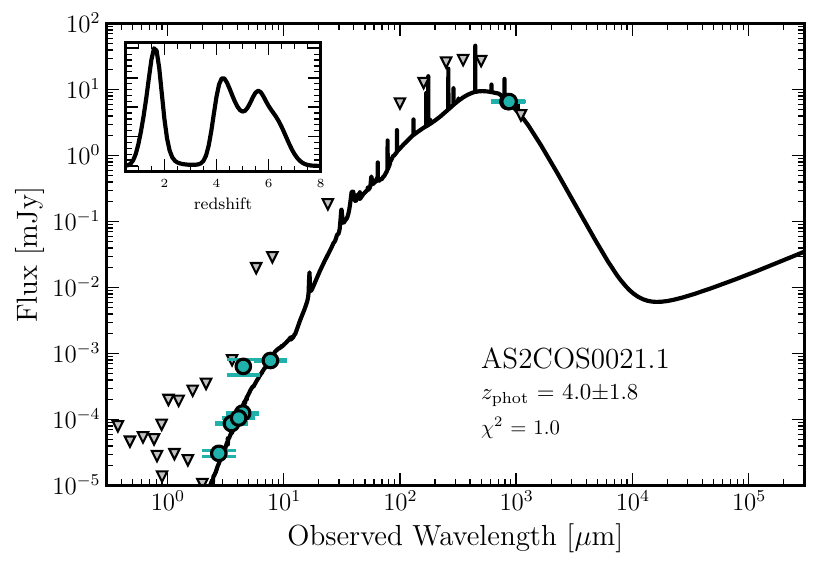}
    \includegraphics[scale=0.33]{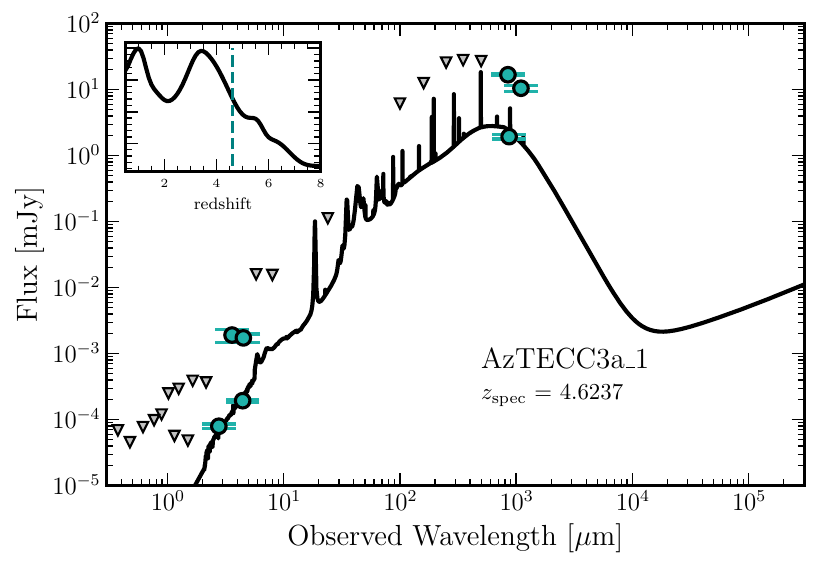}\includegraphics[scale=0.33]{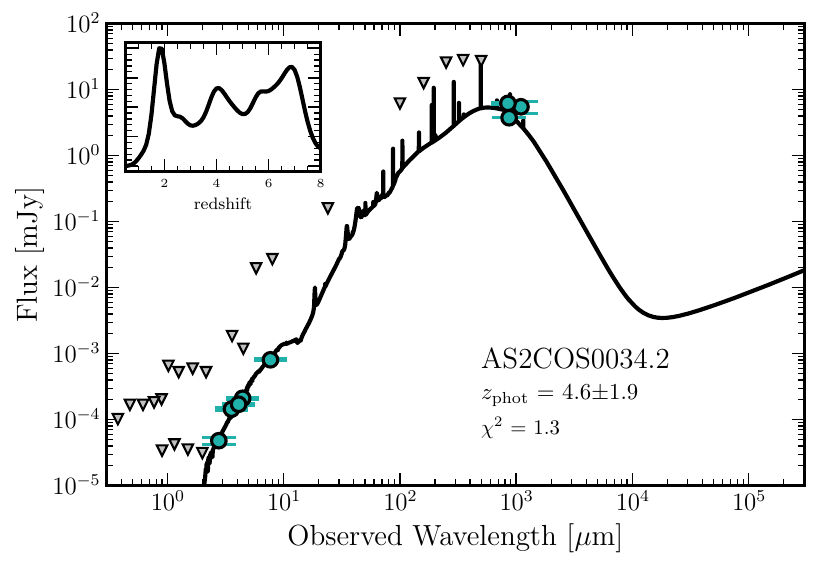}\includegraphics[scale=0.33]{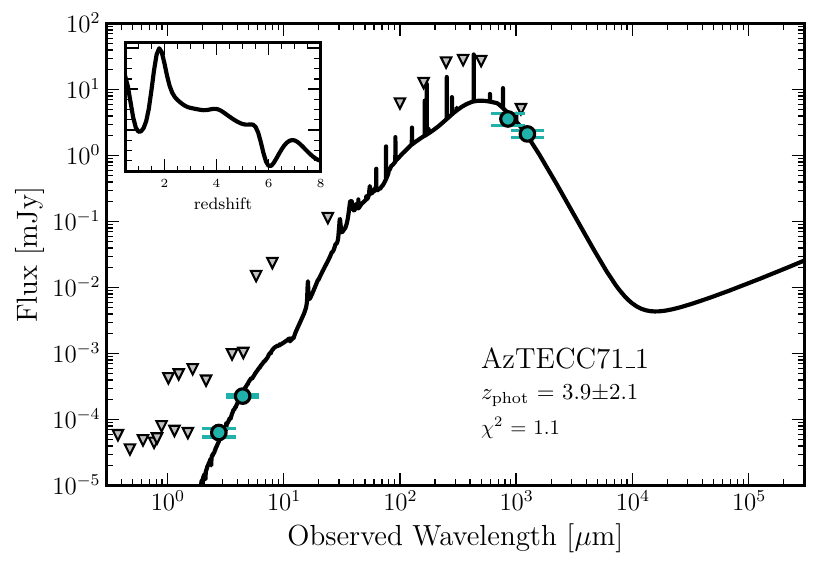}\includegraphics[scale=0.33]{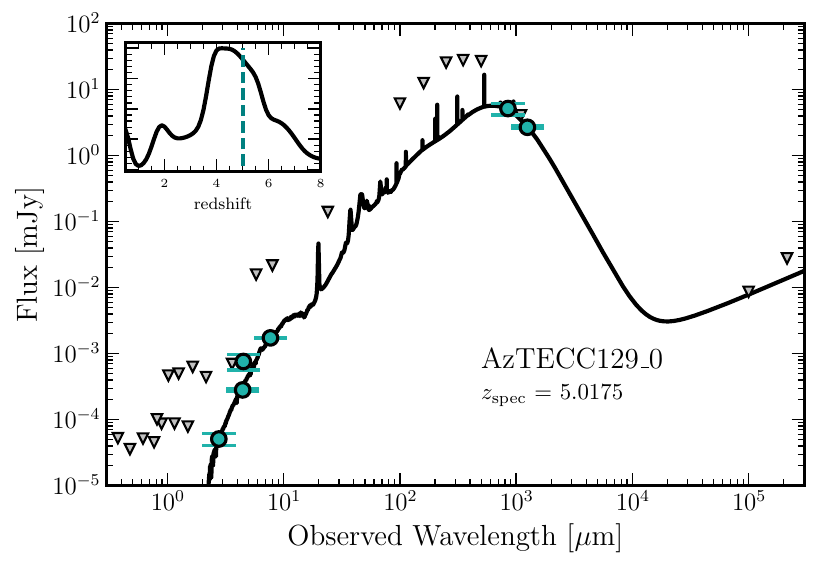}
    \includegraphics[scale=0.33]{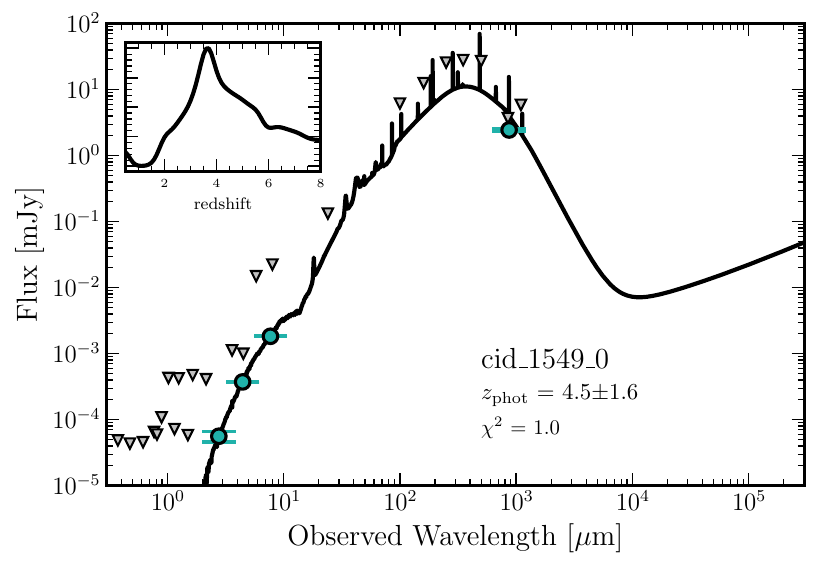}\includegraphics[scale=0.33]{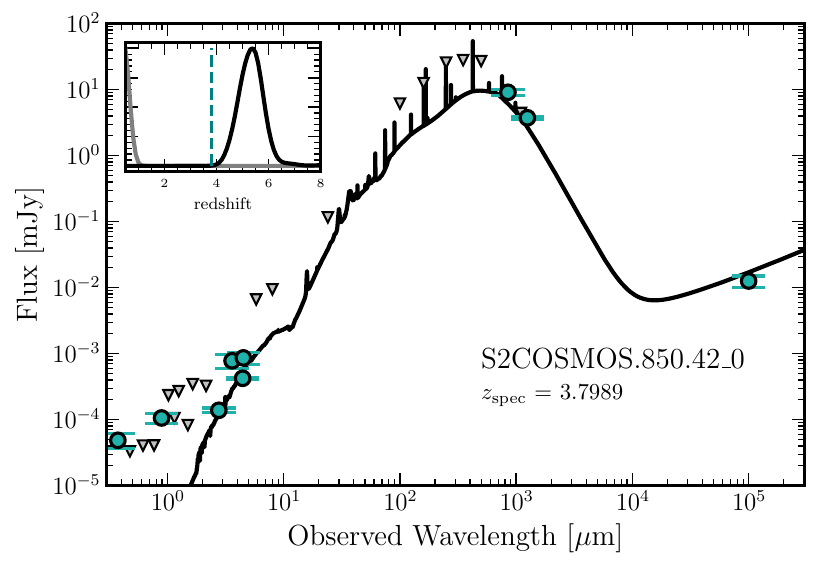}\includegraphics[scale=0.33]{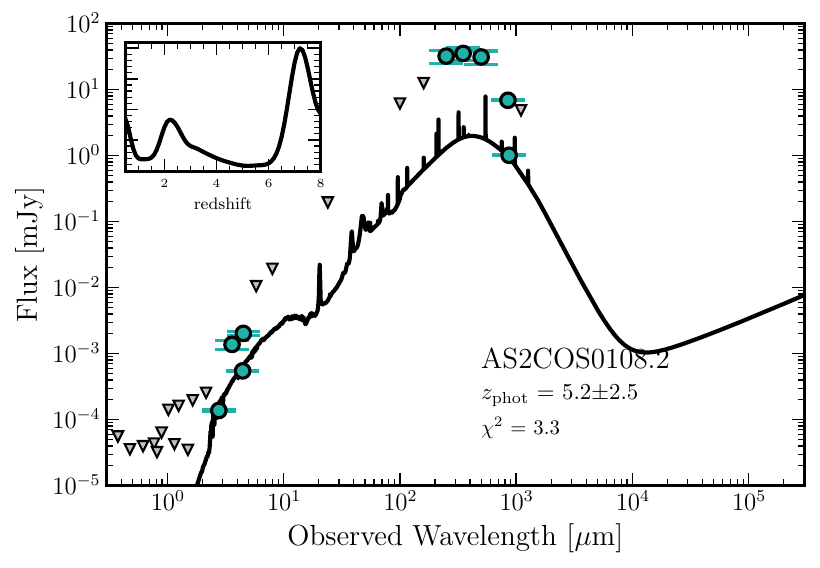}\includegraphics[scale=0.33]{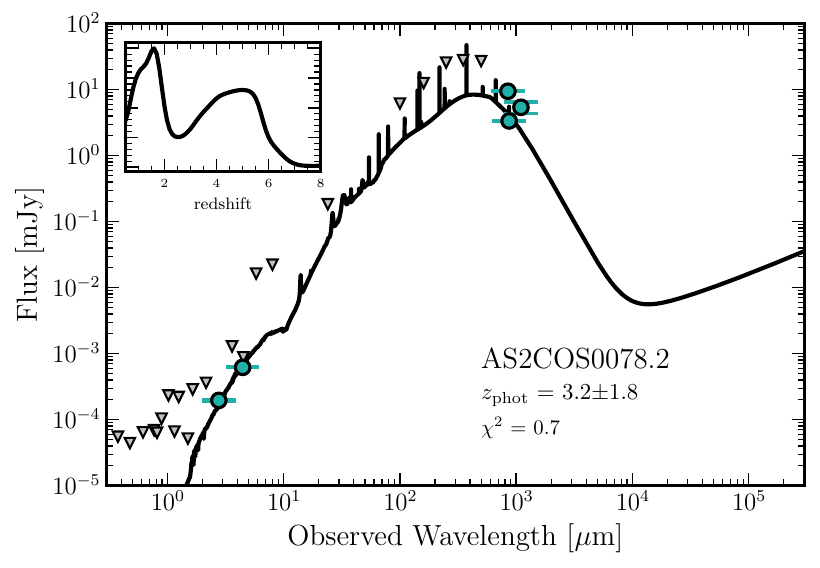}
    \includegraphics[scale=0.33]{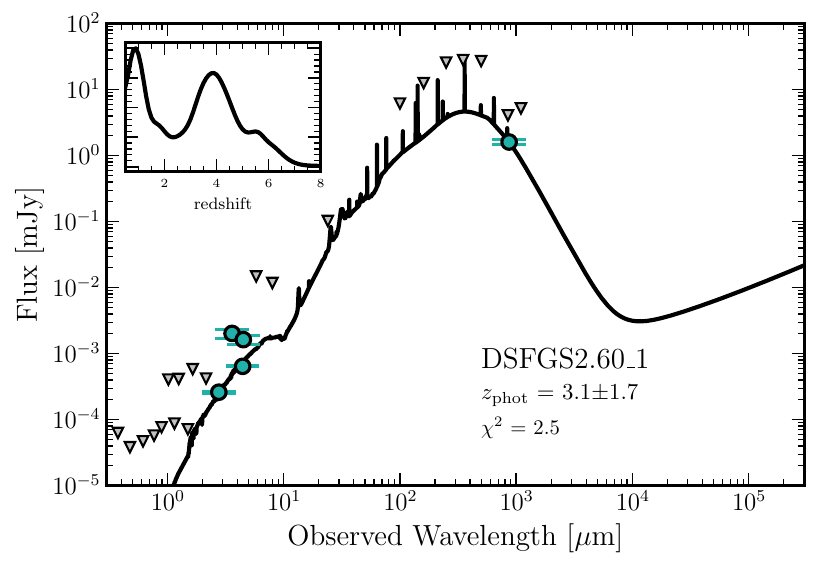}\includegraphics[scale=0.33]{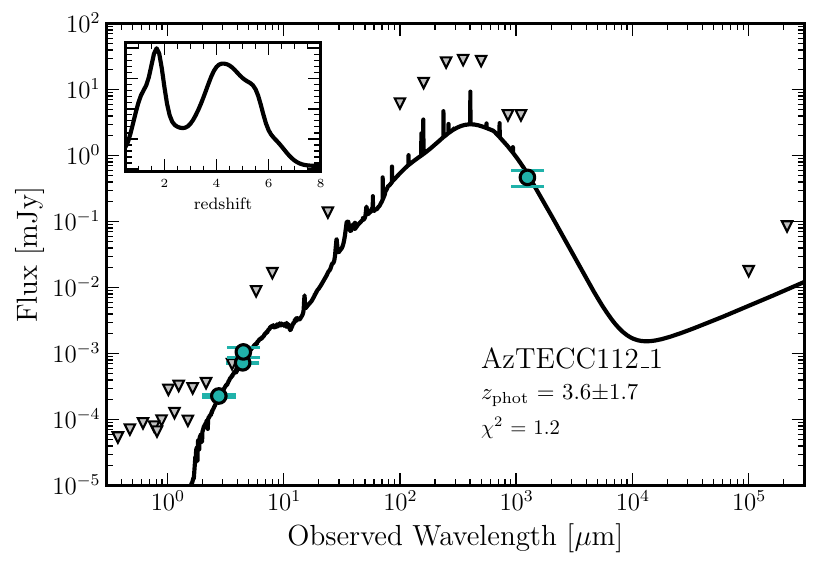}\includegraphics[scale=0.33]{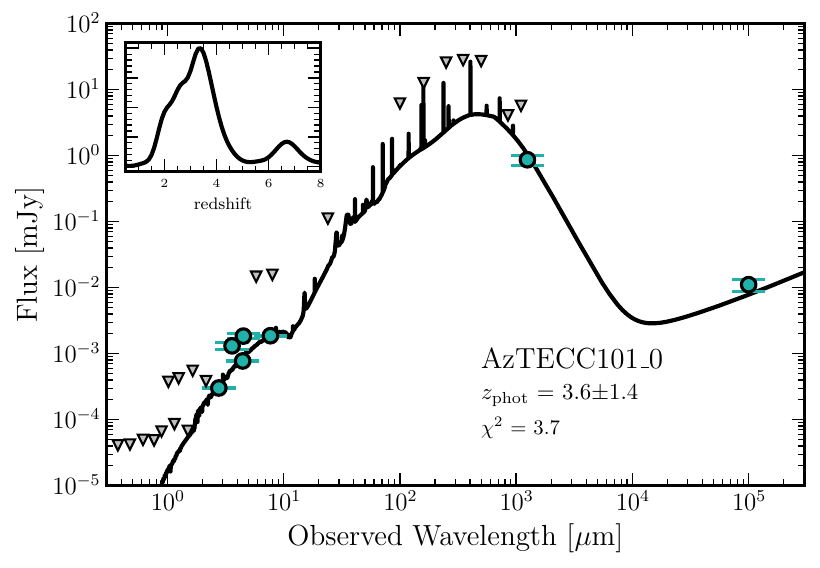}\includegraphics[scale=0.33]{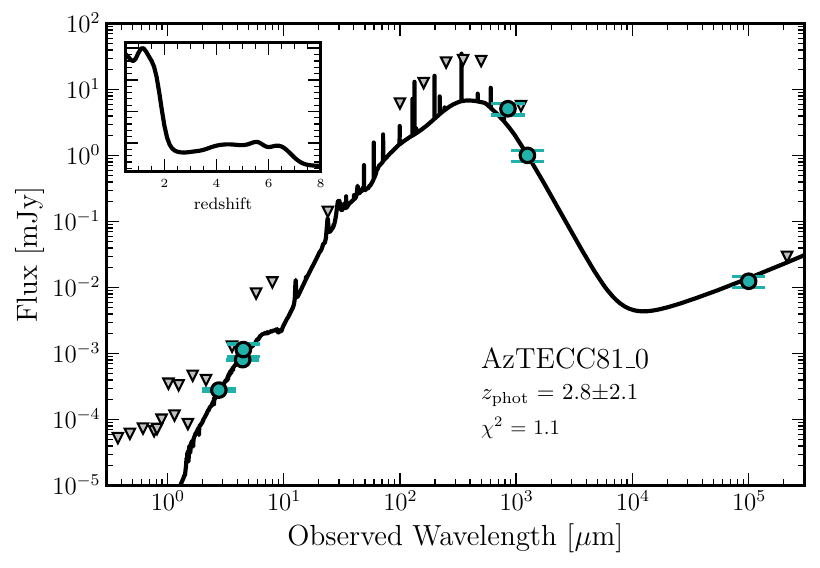}
    \caption{SEDs and associated redshift probability distributions (inset) of the 16 faintest (F444W\,$>24$\,mag) F150W-dropout galaxies from SCUBADive. All data (teal circles) represent $>3.5\sigma$ detections and downward triangles depict $3.5\sigma$ upper limits. Grey PDFs (e.g. in the cases of AS2COS0034.1 and S2COSMOS.850.42.0) indicate misidentifications in \cigale\ resulting from the optical/NIR-only fit. In these cases we adopt the $z_{\rm phot}$ and PDF from the panchromatic SED fit if no $z_{\rm spec}$ exists. Teal dashed lines in the inset plot indicate spectroscopic redshift solutions when available.} 
    \label{fig:seds}
\end{figure*}

SEDs of the 16 faintest targets (F444W\,$>24$\,mag) are shown in Figure \ref{fig:seds} along with their associated redshift probability distribution functions (PDFs) in the upper left inset of each panel. The SEDs come from the optical-NIR only run of \cigale\, which explains the failure of some to fully capture the FIR data. Spectroscopic redshifts from the literature exist for 11/60 of the F150W-dropouts. Seven have $z_{\rm spec}>3.5$, in line with what we would expect for the selection of these galaxies, while four have $z_{\rm spec}\sim2.5$. Even with the additional \jwst\ observations, the heavily obscured nature of these galaxies makes SED fitting difficult and leads to broad and often double peaked photometric redshift distributions. This is seen most noticeably in cases where only a couple bands of photometry exist to constrain the entire rest-frame UV/optical fit. It is encouraging to see many of our redshift estimates converge to high-$z$ solutions (40\% with $z>4$ and 60\% with $z>3.5$) given this was our expectation in selecting for F150W-dropout galaxies, but spectroscopic redshift confirmation remains a top priority as we work to refine these studies. Importantly, the inclusion of \jwst\ data (resulting in smaller redshift uncertainties than pre-\jwst\ works) now allows for more efficient design of spectroscopic follow-up programs. 

\section{Results}\label{sec:results}
\subsection{Finding High Redshift DSFGs while ``SCUBA-Diving''}
\label{subsec:scubadiving}
In total, 60 F150W-dropout galaxies are identified in the SCUBADive sample. Of these 60, 16 ($\sim$27\%) are of particular interest due to their similar photometric properties to AZTECC71, and thus prime targets in our search for the most distant and dusty galaxies. They share faint F444W magnitudes ($>24$\,mag), high photometric redshift solutions ($z_{\rm phot}>3.5$), and have no recorded counterparts in the COSMOS2020 catalog \citep{Weaver2022ApJS..258...11W}. Source names are adopted from those listed in the ALMA Archive to allow for simpler data acquisition, with the exception of those in AS2COSMOS \citep{Simpson2020MNRAS.495.3409S}. Figure \ref{fig:cutouts} shows $4\arcsec\times4\arcsec$ cutouts from COSMOS-Web \jwst\ NIRCam imaging and a combined RGB image (F115W$+$F150W/F277W/F444W). MIRI F770W imaging is also shown when coverage is available. Additionally, we show cutouts from ALMA and VLA with orange and yellow contours depicting their respective emission at $3\sigma$, $4\sigma$, $5\sigma$, and $10\sigma$ levels. Lastly, the light blue line displays the custom aperture created by \texttt{diver}. Table \ref{tab:jwst_data} lists fluxes from the available \jwst\ and \textit{Spitzer} IRAC data and Table \ref{tab:fir_data} lists those from the FIR/mm/radio data.

\subsection{Derived Physical Properties}
\label{subsec:derived_properties}
A list of derived physical properties is reported in Table \ref{tab:properties}. As is expected for galaxies with significant (sub)millimeter flux densities, all of our sources are brightly emitting in the FIR, with total FIR luminosities comparable to that of ULIRGs ($L_{\rm FIR}\approx10^{12}-10^{13}$\,L$_\odot$). The high FIR luminosities of the galaxies are consistent with the conclusion they are undergoing intense bursts of star formation with SFR estimates ranging from $300-1200$\,M$_\odot$\,yr$^{-1}$. The low resolution ALMA data allow us to be confident in our total flux measurements (i.e. that we are not resolving out flux), but higher resolution millimeter observations will shed light on the exact location of where this obscured star formation is taking place in the galaxies.

Studies of heavily obscured galaxies often struggle to produce well-constrained estimates of stellar mass due to a lack of photometric detections to constrain the short wavelength regime of their SEDs. Now, with sensitive \jwst\ observations from the COSMOS-Web survey we are able to improve the SED modeling with the inclusion of available imaging in the F277W, F444W, and occasional F770W filters. While this may seem trivial in the broader context of multiwavelength galaxy studies, for ``dark'' SMGs/DSFGs this is a significant improvement when many heavily obscured galaxies may only have low-resolution IRAC observations to constrain the entire unobscured stellar component. 
For the F150W-dropout sample we find the following average properties: $\langle z \rangle = 3.7^{+0.9}_{-0.7}$, $\langle$SFR$\rangle=450^{+920}_{-320}$\,\mdot\,yr$^{-1}$, $\langle$log(\mstar)$\rangle=11.2^{+0.5}_{-0.6}$\,\mdot, $\langle$log(M$_{\rm dust}$)$\rangle=9.3\pm0.5$\,\mdot, and $\langle A_V \rangle = 4\pm1$. As desired, the dropout selection produces a significantly higher median redshift compared to other DSFG/SMG samples that do not employ this type of cut (e.g., $\langle z \rangle =2.5\pm0.2$; \citealt{Simpson2014ApJ...788..125S} and $\langle z \rangle =2.61\pm0.08$; \citealt{Dudzevi2020MNRAS.494.3828D}). Our findings are in line with results from the radio-selected sample of \cite{Gentile2025arXiv250300112G} ($\langle z \rangle =3.6\pm0.8$) which employs a F150W$>26.1$\,mag cut to require that sources would not be detected at the $3\sigma$ level in H-band in COSMOS2020. Using high resolution radio observations rather than ALMA to locate the correct optical/NIR counterparts has proven to be effective thanks to the large area radio surveys available \citep{Talia2021ApJ...909...23T}, though we note 40\% of F150W-dropouts do not have radio counterparts, highlighting the importance of FIR/mm data in identifying \textit{all} dust-obscured galaxies at early cosmic times.




\begin{deluxetable*}{lCCCCR}
\tablecaption{Derived Properties of F150W-dropout DSFGs \label{tab:properties}}
\tablehead{
\colhead{Name}&
\colhead{Redshift}&
\colhead{SFR}&
\colhead{M$_{\rm dust}$}&
\colhead{M$_\star$}&
\colhead{\av} \\
\colhead{} & \colhead{} & \colhead{[M$_\odot$\,yr$^{-1}$]} & \colhead{[$10^{9}$\,M$_\odot$]} & \colhead{[$10^{10}$\,M$_\odot$]} 
}
\startdata
AS2COS0034.1 & 6.5\pm0.8 & 680\pm40 & 3.6\pm0.3 & 6.6\pm2.1 & 6.1\pm0.3 \\
S2COSMOS.850.50\_1 & 5.8500 & 930\pm220 & 0.7\pm0.2 & 18\pm16 & 7.2\pm0.5 \\
AS2COS0143.1 & 3.9\pm2.1 & 570\pm60 & 3.1\pm0.3 & 8.9\pm4.7 & 7.4\pm0.5 \\
AS2COS0021.1 & 4.0\pm1.8 & 590\pm70 & 3.3\pm0.3 & 9.7\pm1.2 & 6.99\pm0.04 \\
AzTECC3a\_1 & 4.6237 & 260\pm70 & 1.2\pm0.3 & 3.4\pm1.6 & 4.1\pm0.3 \\
AS2COS0034.2 & 4.6\pm1.9 & 460\pm80 & 2.1\pm0.4 & 6.1\pm1.1 & 5.0\pm0.1 \\
AzTECC71\_1 & 3.9\pm2.1 & 430\pm90 & 2.1\pm0.4 & 8.5\pm4.9 & 6.1\pm0.4 \\
AzTECC129\_0 & 5.0175 & 470\pm140 & 2.9\pm0.5 & 26\pm21 & 5.4\pm0.5 \\
cid\_1549\_0 & 4.5\pm1.6 & 810\pm280 & 0.8\pm0.5 & 28\pm13 & 6.1\pm0.3 \\
S2COSMOS.850.42.0 & 3.7989 & 590\pm50 & 3.4\pm0.3 & 1.1\pm0.5 & 5.6\pm0.5 \\
ASCOS00108.2 & 5.2\pm2.5 & 110\pm60 & 0.2\pm0.1 & 2.2\pm0.9 & 3.7\pm0.5 \\
AS2COS0078.2 & 3.2\pm1.8 & 420\pm70 & 1.9\pm0.3 & 7.5\pm3.7 & 5.5\pm0.5 \\
DSFG2.60\_1 & 3.1\pm1.7 & 170\pm50 & 0.7\pm0.1 & 3.0\pm1.2 & 4.1\pm0.3 \\
AzTECC112\_1 & 3.6\pm1.7 & 140\pm60 & 0.6\pm0.2 & 7.2\pm4.1 & 4.1\pm0.4 \\
AzTECC101\_0 & 3.6\pm1.4 & 230\pm30 & 1.0\pm0.3 & 2.6\pm0.7 & 3.0\pm0.2 \\
AzTECC81\_0 & 2.8\pm2.1 & 230\pm30 & 1.3\pm0.2 & 4.7\pm1.7 & 4.9\pm0.3 \\
\enddata
\tablecomments{Derived physical properties from \texttt{CIGALE} for the faintest F150W-dropouts found via SCUBA-diving in COSMOS-Web. Redshift values without uncertainties represent spectroscopically confirmed redshifts \citep{Casey2019ApJ88755C, Chen2022ApJ...929..159C}.}
\end{deluxetable*}


\section{Discussion}\label{sec:discussion}

\subsection{Leveraging \jwst\ Imaging to Select the Highest Redshift DSFGs}
The current area coverage of ALMA programs overlapping with the COSMOS Survey field of view is only $\sim5\%$ of the total survey area and thus the major limiting factor in our search for more dust-obscured galaxies utilizing the selection technique described here. Future large area, blind surveys with instruments like TolTEC on the Large Millimeter Telescope (LMT; \citealt{Hughes2020SPIE11445E..22H}) or the New IRAM KID Array 2 (NIKA2; \citealt{Bing2023A&A...677A..66B}) on the IRAM 30\,m telescope can efficiently map larger portions of the sky to further increase our sample size of $z>4$ dusty galaxies. While it may be possible to find candidate high-$z$ dusty galaxies with \jwst\ photometry alone, there will inevitably be difficulty in selecting clean samples that are not contaminated by other extremely red objects (EROs; \citealt{Barro2024ApJ...963..128B,Sun2025ApJ...987...60S}), namely: $z>3$ quiescent galaxies (QGs; \citealt{Carnall2020MNRAS.496..695C,Long2023arXiv230504662L,Antwi-Danso2023arXiv230709590A,Barrufet2025MNRAS.tmp...22B}) and $z>4$ little red dots (LRDs; \citealt{Labbe2023Natur.616..266L,Barro2023arXiv230514418B,Kokorev2024ApJ...968...38K}). A color-magnitude selection wedge is offered in \citetalias{McKinney2025ApJ...979..229M} (Figure 11 therein) to select all $z>3$ SCUBADive sources and a clear, albeit unsurprising, trend emerges: DSFGs on average get redder in F277W-F444W color and get fainter in F444W with increasing redshift. \jwst\ imaging might aid in selecting the highest redshift candidates from existing (sub)millimeter surveys when spectroscopic observations are unavailable, but it is clear that DSFGs cannot be selected from \jwst\ imaging alone. 

\subsubsection{Potential Confusion with Little Red Dots?}
A major topic of discussion in the literature as of late is the emergence of a new population of galaxies being readily discovered by \jwst; LRDs. Given the similar red colors between LRDs and the dusty galaxies investigated in this work, we ask, do any of our sources qualify as LRDs? Alternatively, how many LRDs might actually be compact, high-$z$ DSFGs? We find that 85\% of our sources abide by the F277W$-$F444W\,$>1$ criteria described in \cite{Labbe2025ApJ...978...92L}, but none possess the significant unobscured rest-UV emission to reproduce the moderate blue continuum necessary for the color selections in the shorter wavelength filters (a predictable result given that they are F150W-dropouts). Similarly, 15\% of our sources have F277W$-$F444W\,$>1.5$ as required by \cite{Akins2024arXiv240610341A}, but none satisfy the compactness criteria in F444W imaging. Our findings, or rather lack thereof, agree with those from \cite{Labbe2025ApJ...978...92L}: comparably red, star-forming samples may satisfy the color selections for LRDs, but are extended and detected by ALMA in nearly all cases. 



\subsubsection{Testing Source Detectability with UNCOVER}
A question that is often brought up when discussing UV-optical ``dark'' galaxies is whether or not they would be detected at shorter wavelengths simply if deeper data were available. How much deeper would we have to go for our sample of galaxies to lose their dark label? To test this, we compare to the UNCOVER survey \citep{Bezanson2024ApJ...974...92B} which has some of the deepest publicly available \jwst\ NIRCam data and associated high level science products. We first re-do aperture photometry for the F150W-dropouts, this time using the aperture size implemented in the (non-variable aperture) UNCOVER photometric catalog (0\arcsec.32) rather than the \diver\ custom apertures. Next, we re-run \cigale\ with the aperture-matched photometry to obtain predicted model flux densities from SED fitting for the short wavelength filters, F115W and F150W. Finally, we compare these model flux densities to the UNCOVER depths to test if we would detect these galaxies given the deeper data. Assuming the model flux densities are accurate, all of the F150W-dropout galaxies would be detectable in F150W if we had depths comparable to the UNCOVER survey (F150W $5\sigma$ depth $=30.18$). However, even with this impressive dataset, roughly 40\% would remain undetected in F115W ($5\sigma$ depth $=30.05$), demonstrating the incredibly obscured nature of these sources. Achieving these depths in UNCOVER required six hours of integration time for both F115W and F150W and obtaining such pristine observations over wide areas to further reveal the rest-frame UV emission of rare DSFGs may prove a difficult ask. That said, we have started to see the potential power of pure parallel observations with \jwst\ 
which can result in very deep imaging if the primary program requests many hours for its independent science goal. Searching for DSFGs in blank field parallel observations, e.g., the PANORAMIC Survey (PID 2514: PIs Williams \& Oesch; \citealt{Williams2025ApJ...979..140W}) may yield unique sources that give us the best view of unobscured star-formation in $z>3$ DSFGs yet. Efficient ALMA continuum snapshot surveys targeting the recovered candidates could then quickly confirm their obscured nature.

\subsection{Comparing ``Dark'' Galaxies}
\label{subsec:nirdark_compare}
The search for heavily dust-obscured star-forming galaxies at $z>3$ has rapidly expanded within the last few years, especially now that deep \jwst\ observations are able to trace previously undetected stellar components for many of these systems. As sample sizes of ``dark'' galaxies continue to increase, we must take care in comparing the sources found via the various selection techniques and survey depths across the literature. Attempts to place our discoveries into neatly defined categories may not be the most useful route as we move beyond ``stamp collecting'' into larger comprehensive population studies.
It may be that these methods are selecting different sub-populations of DSFGs. On the other hand, the searches may recover galaxies with essentially the same formation histories and physical properties, but different dust geometries and dust covering fractions \citep{Cochrane2024ApJ...961...37C}. Distinguishing between these scenarios will require larger sample sizes, resolution-matched (sub)millimeter data to pinpoint dust emission, and spectroscopic observations to constrain redshifts and star-formation histories among other key properties. Additionally, more robust constraints on stellar mass would allow for cleaner mass-selected samples. For now, we use the data available to compare the F150W-dropouts of this work with galaxies selected via other popular ``dark'' selection criteria.

\begin{figure}
    \centering
    \includegraphics[width=\columnwidth]{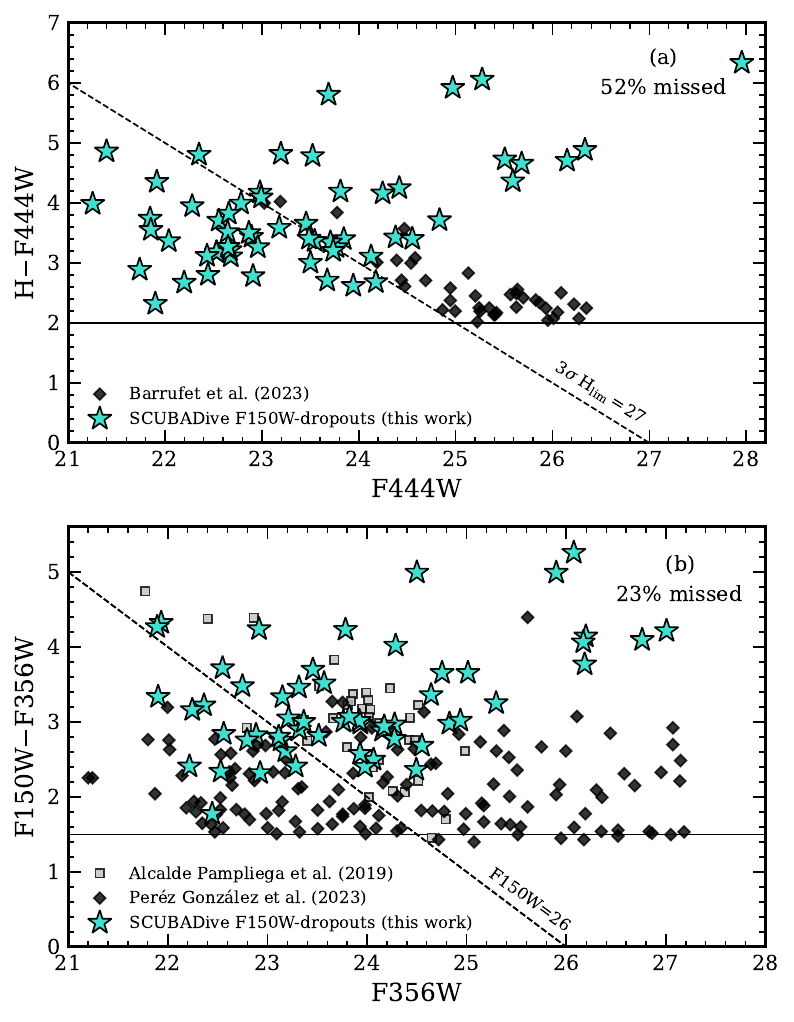}
    \caption{Significant fractions of $z>3$ dust-obscured galaxies can be missed when relying on rest-frame optical color-magnitude selections alone. Here are two color-magnitude selections for ``dark'' galaxies from the literature and the percentage of our sources (cyan stars) missed by these criteria. \textit{(a)} Selection from \cite{Barrufet2023MNRAS.522..449B} which makes use of the \hst\ H-band (F160W) and \jwst\ F444W filters. \textit{(b)} Selection from \cite{Perez-Gonzalez2023ApJ...946L..16P} which makes use of two \jwst\ filters, F150W and F356W. As our sources are not formally detected in F150W and we do not have F356W data, we leverage our best-fit model convolved with the \jwst\ filters in order to infer their F150W-F356W colors.}
    \label{fig:compare_nirdark_selections}
\end{figure}

In Figure \ref{fig:compare_nirdark_selections} we examine whether our F150W-dropout galaxies would be selected with popular color-magnitude selections from other works and where the sample falls in these parameter spaces. For filters that are either undetected or unavailable in the COSMOS/COSMOS-Web surveys, we interpolate the galaxy model SEDs from \cigale\ to retrieve fluxes, and thus magnitudes, at the necessary wavelengths. Panel (a) shows the color cut implemented by \cite{Barrufet2023MNRAS.522..449B} ($H-F444W>2.0$ and $H>27$) which results in a total of 30 ``\hst-dark'' galaxies found within the CEERS survey \citep{Finkelstein2023ApJ...946L..13F}. This selection was implemented specifically to identify galaxies at $z>3$ with $A_v>2.0$. Despite over 90\% of our sources sharing these desired properties according to SED fitting results, a little under half are recovered with the H-band/F444W selection. This is where survey depth and available wavelength coverage come into play. The F150W-dropout galaxies are almost entirely (54/60 objects) H-band dark in the COSMOS survey, but the H-band data available are from ground-based VISTA observations and are 2--3 orders of magnitude shallower than the \hst\ F160W observations in the CEERS field. This suggests caution should be taken when applying stringent color-magnitude cuts defined by survey depth as they can result in sample incompleteness of high redshift dusty galaxies. The efforts described in this work to select $z>4$ DSFGs still rely on a somewhat arbitrary cut anchored to the F150W depth of COSMOS-Web by definition, though considering the vastly improved sensitivity of \jwst\ over previous studies we presume that the available \jwst\ F150W imaging should nominally detect all classic SMGs/DSFGs at $z<3-4$. Selections such as these are a perfectly fair place to start when most surveys only have imaging and continuum data to work with, but hopefully we can begin to move towards more meaningful, physically motivated identifiers (beyond their dropout filter) of the galaxies that compose the dust-obscured Universe.

Panel (b) shows the selection from \cite{Perez-Gonzalez2023ApJ...946L..16P} (including the sample from \citealt{AlcaldePampliega2019ApJ...876..135A}) which makes use of two slightly bluer filters ($F150W-F356W>1.5$ and $F356W<27.5$) in the CEERS survey. They find 138 galaxies that fit this criteria. The dashed diagonal line denotes the differentiation between \hst-faint and \hst-dark (F150W$>26$\,mag) galaxies. 77\% of our F150W-dropouts are selected as \hst-dark galaxies with this criteria, though we note they have redder colors than the majority of the literature sample. Works like this and others have demonstrated the utility of color-magnitude selections in identifying the reddest objects ever observed, many of which appear to be dust-obscured galaxies. However, they also do well to clearly illustrate one major stipulation: multiple distinct galaxy populations (i.e. dusty galaxies, quiescent/post-starburst galaxies, and $z>6$ star-forming galaxies) share similar red colors and can all be subsumed under the vague ``dark'' label (see also \cite{Barrufet2025MNRAS.tmp...22B} for a spectroscopic study). Given this understanding, we stress the inclusion of FIR/mm data to unambiguously confirm the presence of dust as we focus our analysis on uncovering and characterizing obscured star formation at early times. Relying on selections based on \jwst\ NIRCam imaging alone will result in incomplete samples of dusty galaxies. 

\subsection{Are ``Dark'' Galaxies Extreme?}
\label{sub:are_dark_gals_extreme}
\begin{figure}
    \centering
    \includegraphics[width=\columnwidth]{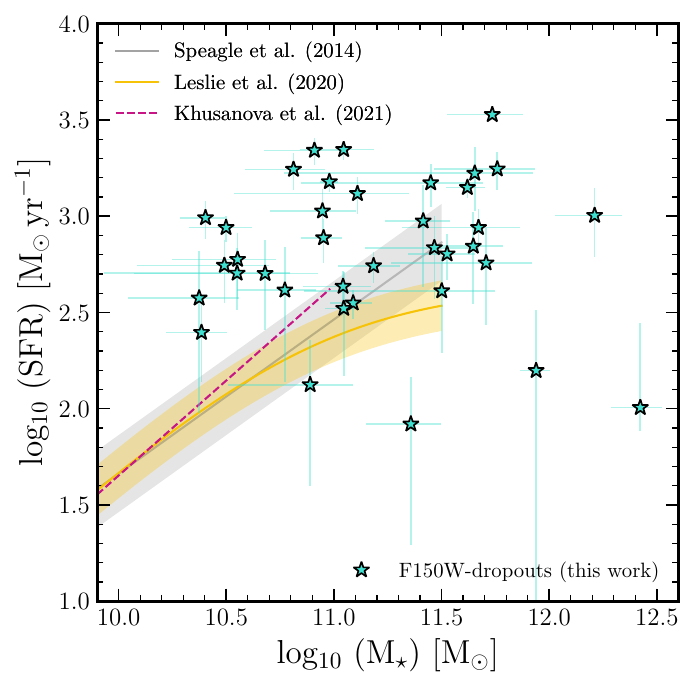}
    \caption{SFR versus \mstar\ main sequence at $z\sim4$ defined by \cite{Speagle2014ApJS..214...15S} (grey line where shaded region depicts $\pm0.2$\,dex), \cite{Leslie2020ApJ...899...58L} (yellow line where shaded region shows the range between $3<z<6$), and \cite{Khusanova2021A&A...649A.152K} (magenta dashed line). Here we only plot the 37/60 sources with $z_{\rm phot}$ or $z_{\rm spec}>3.5$ for fair comparison. We see decent scatter about the MS, suggesting these ``dark'' galaxies are fairly typical for their redshifts and physical properties, though not necessarily members of a uniform population given the spread in parameter space.}
    \label{fig:SFMS}
\end{figure}

Much of the ongoing conversation around ``dark'', dust-obscured galaxies centers on whether or not they are extreme sources compared to the broader star-forming galaxy population at early times. In order to contextualize what ``extreme'' is at these epochs, we look to the main sequence (MS; \citealt{Noeske2007ApJ...660L..43N}) of star-forming galaxies and examine how the F150W-dropout sub-population compares given our current observations. In Figure \ref{fig:SFMS}, we show our $z>3.5$ DSFGs in relation to the observed correlation between \mstar\ and SFR in galaxies, as parameterized by \cite{Speagle2014ApJS..214...15S}, \cite{Leslie2020ApJ...899...58L}, and \cite{Khusanova2021A&A...649A.152K}.

We include three parameterizations of the MS, two which maintain linear forms and one which exhibits a turnover, or flattening, at high stellar masses (\mstar$>10^{10.5}$\mdot). We do this since a consensus has not yet been reached in the literature on the main sequence's proper characterization. In \cite{Speagle2014ApJS..214...15S}, the MS is derived from a compilation of 25 studies from the literature. \cite{Leslie2020ApJ...899...58L} opt for a 3\,GHz radio-selected sample to construct the MS for a dust-unbiased view of SFR. Finally, the MS determined by \cite{Khusanova2021A&A...649A.152K} is based on ``normal'' star-forming galaxies at $z\sim4.5$ from the ALMA Large Program to INvestigate [C\textsc{ii}] at Early times (ALPINE; \citealt{Faisst2020ApJS..247...61F}), though the UV-selection utilized means by design the UV/optically-dark galaxies are not included. Depending on which form is used in our analysis, different conclusions regarding the nature of these galaxies may be reached, though it is also important to note that none of the parameterizations are particularly well constrained beyond \mstar$>10^{11}$\mdot\ where the majority of our sources lie. 

Nearly half (48\%) of $z_{\rm phot}>3.5$ sources lie more than $3\times$ above the \cite{Leslie2020ApJ...899...58L} $z\sim4$ MS track, indicative of a starburst event (though not necessarily extreme) for its epoch and mass regime \citep{Elbaz2018A&A...616A.110E}. Fewer fall into this category if we instead look at their distance from the linear forms as 35\% and 27\% are considered starbursts compared to the \cite{Speagle2014ApJS..214...15S} and \cite{Khusanova2021A&A...649A.152K} relations, respectively, if we extrapolate out to higher masses. It appears a few of the F150W-dropout DSFGs may have slightly elevated SFRs, but none look to be particularly extreme galaxies and are consistent with the high-mass end of the MS. 
Regardless of the adopted MS, what seems apparent based on their spread across the \mstar-SFR parameter space is that $z>3.5$ DSFGs do not belong to a uniform population just as it has been suggested for lower redshift SMGs/DSFGs around cosmic noon \citep{Hayward2012MNRAS.424..951H,DaCunha2015ApJ806110D,Gillman2024arXiv240603544G}. How these galaxies might be subdivided by morphology, environment, triggering mechanism, dust geometry, etc., will be important to address with future observations.

We note our full sample has quite high stellar mass estimates (62\% with \mstar$>10^{11}$\mdot) even after some accounting for model discrepancies was done in \citetalias{McKinney2025ApJ...979..229M} (see Section 7.4 therein). This is not implausible, however. Given what is understood about the relationship between dust-obscured star formation and stellar mass, these galaxies are consistent with the accepted trend of increased obscured fraction at increasingly high mass \citep{Whitaker2017ApJ...850..208W,Zimmerman2024ApJ...973..146Z}. 
The stellar masses also fall in line with expected dust-to-stellar mass ratios of main sequence and starburst galaxies \citep{Donevski2020A&A...644A.144D}. 
Ultimately, the very nature of heavily obscured galaxies means they lack sufficient detections to robustly constrain the rest-frame UV/optical portion of the SED and we severely lack near-IR/mid-IR observations to constrain the older stellar population. In the most extreme cases, just two or three robust detections inform the model fitting for the entire UV/optical/IR regime. Finally, we note that while we cannot completely rule out the presence of AGN in our sample, which could cause overestimates of stellar mass, none exhibit signs of AGN dominated light morphologically. Future rest-frame mid-IR \jwst\ observations will be necessary to constrain stellar masses and reveal clearer signs of AGN through SED fitting, e.g. observations of hot torus dust. Without the appropriate data to meaningfully probe potential AGN contribution, we refrain from any further speculation at this time. 



\subsection{Contribution of ``Dark'' Galaxies to the Stellar Mass Function}
Defined as the number density of galaxies as a function of their stellar mass at a given redshift, $\Phi(M,z)$, the galaxy stellar mass function (SMF) allows us to quantify the assembly and growth of baryons and track early galaxy evolution. Efforts from both sides of theory \citep{Long2023ApJ...953...11L} and observations \citep{Gottumukkala2024MNRAS.530..966G} have worked to assess how much the high-mass end of the SMF may be affected due to the historical reliance on UV/optical tracers to select galaxy populations which miss many massive and dust-obscured star-forming galaxies.

\begin{figure}
    \centering
    \includegraphics[width=\columnwidth]{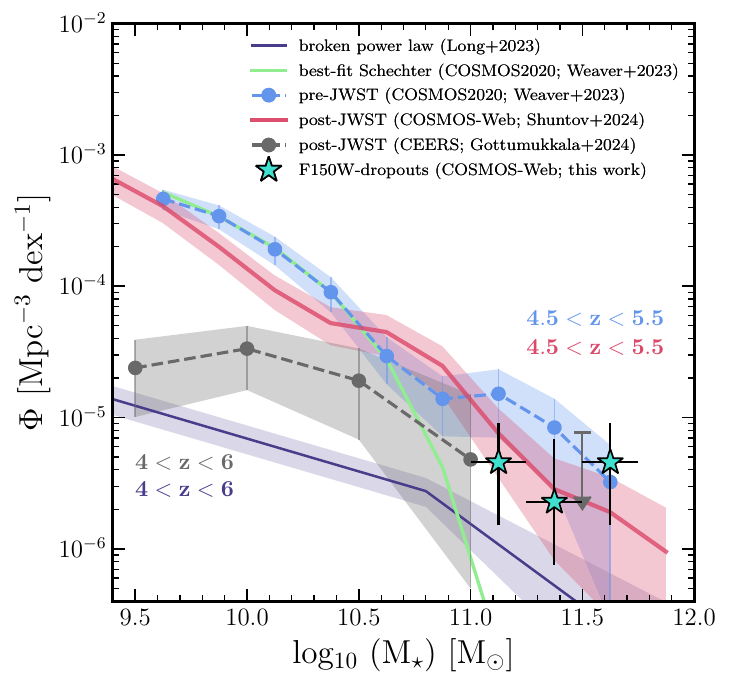}
    \caption{The stellar mass function of high-$z$ galaxies pre-\jwst\ (blue and green lines; \citealt{Weaver2023A&A...677A.184W}) and post-\jwst\ (grey line; \citealt{Gottumukkala2024MNRAS.530..966G} and red line; \citealt{Shuntov2025A&A...695A..20S}). Additionally, a dust-obscured SMF derived explicitly from IR-luminous galaxies is shown \citep[purple line;][]{Long2023ApJ...953...11L}. Cyan stars indicate the potential contribution from the highest mass F150W-dropouts presented in this work that were missed in the COSMOS2020 catalog due to dust obscuration. Strikingly, around 60\% of the galaxy population could have been missed at the highest mass regime (log$_{10}$(\mstar/\mdot)$>11.5$) assuming the redshift and stellar mass estimates are indeed accurate.}
    \label{fig:SMF}
\end{figure}

Figure \ref{fig:SMF} displays three SMFs; one determined from all galaxies recorded in the COSMOS2020 catalog in the 1.55\sqdeg\ COSMOS Survey \citep{Weaver2023A&A...677A.184W}, 
another consisting of just massive and dusty galaxies found by \jwst\ in the 35.5\,arcmin$^{2}$ CEERS survey (grey line; \citealt{Gottumukkala2024MNRAS.530..966G}), and the third from the most recent 0.54\,deg$^{2}$ \jwst\ COSMOS-Web Survey and associated COSMOS2025 catalog \citep{Shuntov2025A&A...695A..20S}. Analysis of the CEERS galaxies, selected to be red, optically faint and thus excluded by the COSMOS2020 selection, suggests that up to $\sim25$\% of galaxies are missed by pre-\jwst\ surveys at the high-mass end between $z\sim4-6$. Noticeably, the CEERS sample has smaller stellar masses on average compared to our SCUBADive sample and the derived SMF stops at log$_{10}$(\mstar/\mdot)=11. Taking the SCUBADive stellar mass and photometric redshift estimates at face value, we add the F150W-dropouts to the SMF plot in order to extend this analysis beyond log$_{10}$(\mstar/\mdot)$>11$ and probe the very high-mass end (cyan stars). 

As discussed in \citetalias{McKinney2025ApJ...979..229M}, SCUBADive suffers from sample incompleteness relative to the total available \jwst\ imaging due to the nature of the ALMA archival search employed and lack of uniform ALMA coverage across COSMOS-Web. Only $\sim31$\% of SCUBA-2 sources (219/706) in the COSMOS-Web area have ALMA observations to cross-match to \jwst\ counterparts, mostly from targeted pointings rather than unbiased mosaics. Thus, calculating the number density of F150W-dropouts from the significantly smaller area observed by ALMA would not produce results representative of the true population statistics. To address this, we take a conservative approach and extrapolate the fraction of missing F150W-dropouts to the remaining SCUBA-2 sources, in order to estimate how many F150W-dropouts there may be across the full COSMOS-Web field. For example, between $4.5<z_{\rm phot}<5.5$ we find two F150W-dropouts with stellar mass $11.50<$log$_{10}$(\mstar/\mdot)$<11.75$ not in COSMOS2020. This is $\sim0.7$\% of the current SCUBADive sample. If we take this percentage of the parent sample of 706 SCUBA sources within COSMOS-Web after accounting for the multiplicity fraction of SCUBA-2 sources in SCUBADive ($\times1.32$), we may expect a total of $\sim$six missed F150W-dropout sources in this redshift and mass range. It is with this assumed source count that we then calculate the volume density to compare to the SMF. This process is then repeated for the remaining two mass bins. Horizontal error bars indicate the sizes of the mass bins. Vertical error bars indicate the upper and lower $2\sigma$ confidence intervals as set by Poisson statistics \citep{Gehrels1986ApJ...303..336G}, which we expect to be the most significant source of uncertainty at the high mass end given the small sample sizes.

This examination suggests that $>20$\% of galaxies may have been missed from pre-\jwst\ catalogs and SMFs, similar to the conclusions from \cite{Gottumukkala2024MNRAS.530..966G} for the slightly lower mass population. In our highest mass bin ($11.5<$log$_{10}$(\mstar/\mdot)$<11.75$) this potential missed fraction reaches up to a staggering $\sim60$\% \textemdash\ in line with previous studies which reported nearly half of the most massive galaxies at $4<z<6$ were missing from \hst\ surveys, but identified by \textit{Herschel} \citep{Liu2018ApJ...853..172L}. These estimates lie above predictions of the dust-obscured SMF derived from IR-luminous galaxies only \citep{Long2023ApJ...953...11L}, suggesting their prevalence is greater than previously thought. 


Now with the inclusion of \jwst\ observations, all of our F150W-dropouts are identified in the COSMOS2025 catalog \citep{Shuntov2025arXiv250603243S} thanks to robust F444W detections, so previous concerns of missing galaxies appear to have been mitigated, though the possibility of NIRCam-dark galaxies persists \citep{Perez-Gonzalez2024ApJ...969L..10P,Sun2025arXiv250606418S}. Still, we see how even just a few heavily-obscured galaxies dominate the make-up of the high mass end of the updated SMF (red line in Figure \ref{fig:SMF}). Adding or removing even a single galaxy to any of the bins, for example as redshift and/or mass measurements are updated, has large implications for the shape of the SMF, supporting the critical need for follow-up observations to better constrain the physical properties of massive, dust-obscured galaxies. Spectroscopic programs will be key to unambiguously determine redshifts and mid-IR observations (rest-frame NIR at these redshifts) with \jwst\ MIRI will offer significant improvements to stellar mass estimates \citep{Papovich2023ApJ...949L..18P, Song2023ApJ...958...82S}. Wider area surveys will also be necessary to find the rarest, most massive systems. Some success has been found via large blind mosaics with ALMA (e.g. Ex-MORA; \citealt{Long2024arXiv240814546L} and CHAMPS; Faisst et al. 2025 in prep.; Martinez et al. 2025 in prep.), but a complete snapshot program of all SCUBA-2 sources would be a valuable starting point to first correctly identify all optical/NIR counterparts. In the meantime, we may look to expand SCUBADive to other extragalactic legacy fields beyond COSMOS-Web to find and compile these rare, hidden systems.




\section{Summary and Conclusions}\label{sec:conclusions}
At this time, the true dust content of the early Universe remains in question and new detections of prolific star-formers previously hidden from view have spurred on searches for more of these intriguing systems. Following the \jwst\ identification of AzTECC71, a FIR bright galaxy with $z_{\rm phot}>5$ that had gone undetected at $\lambda_{\rm obs}<4$\,\micron\ before the launch of \jwst\ \citep{McKinney2023ApJ...956...72M}, and the compilation of all SCUBA-2 detected galaxies with ALMA and \jwst\ counterparts in COSMOS-Web (SCUBADive; \citetalias{McKinney2025ApJ...979..229M}), we search for similarly distant and dusty galaxies. Our findings are as follows:

\begin{itemize}

    \item 60 out of 289 SCUBADive galaxies are F150W-dropouts. 16 of these sources are of particular interest for their fainter F444W magnitudes (mag$_{F444W}>24$), lack of COSMOS2020 catalog counterparts, and thus potential to be among the highest redshift dusty galaxies observed so far with \jwst.
     
    \item SED fitting with \cigale\ produces high stellar mass ($\langle$log(\mstar)$\rangle=11.2^{+0.5}_{-0.6}$\,\mdot), SFR ($\langle$SFR$\rangle=450^{+920}_{-320}$\,\mdot\,yr$^{-1}$), and redshift estimates ($\langle z \rangle = 3.7^{+0.9}_{-0.7}$) for the F150W-dropout sample. However, we find that the majority of these galaxies may not be particularly extreme for their presumed epoch, roughly following the extrapolated trends of current SFR-MS tracks.  

    \item Color and/or dropout selections are helpful in identifying the highest redshift galaxy candidates, but relying on \jwst\ imaging alone will result in interlopers and incomplete samples of dusty galaxies without FIR/mm data to robustly confirm their presence. 

    \item Anywhere from $20-60$\% of galaxies may have been missed due to heavy obscuration and thus unintentionally excluded from the high-mass end (\mstar$>10^{11}$\,\mdot) of pre-\jwst\ derived stellar mass functions. Spectroscopic follow-up and mid-IR observations are needed to confirm these galaxies are indeed in the correct redshift and stellar mass bins, but the current results imply revisions not only to the SMF, but also the cosmic SFH and the role of dust at high redshifts, may be required.
    
\end{itemize}

SCUBADive is an ongoing project to study the \jwst\ counterparts of dust-obscured galaxies. In the COSMOS-Web field alone, 487 SCUBA-2 sources (which could end up being $>600$ discrete sources after accounting for multiplicity) remain unaccounted for in that they lack high-resolution ALMA follow-up to precisely pinpoint their optical/NIR emission. Targeted, observationally inexpensive ALMA snapshots to complete the census of SMGs/DSFGs across the entire 0.54\sqdeg\ COSMOS-Web area are a clear first step to continuing the SCUBADive excursion and furthering our understanding of the dust-obscured Universe at early times. In lieu of a targeted campaign, the ongoing A3COSMOS project \citep{Liu2019ApJS..244...40L} will continue to compile ALMA observations from the archive as they become available. Additional spectroscopic follow-up and rest-frame NIR data will significantly improve redshift and stellar mass constraints. Future SCUBADive works will look to analyze morphologies and as well as conduct spatially-resolved SED fitting to elucidate formation scenarios. Finally, we hope to expand SCUBADive to other extragalactic legacy fields in an effort to further assemble the census of dusty and ``dark'' galaxies.

\begin{acknowledgments}
S.M. Manning would like to thank her older brother, Alex, for the many hours spent co-working together, during which much of this manuscript was written. Support for programs JWST-AR-5213 was provided by NASA through a grant from the Space Telescope Science Institute, which is operated by the Associations of Universities for Research in Astronomy, Incorporated, under NASA contract NAS 5-03127. Support for this work was provided by NASA through the NASA Hubble Fellowship grant \#HST-HF2-51484 awarded by the Space Telescope Science Institute, which is operated by the Association of Universities for Research in Astronomy, Inc., for NASA, under contract NAS5-26555. We honor the invaluable labor of the maintenance and clerical staff at our institutions, whose contributions make our scientific discoveries a reality. The University of Massachusetts acknowledges that it was founded and built on the unceded homelands of the Pocumtuc Nation on the land of the Norrwutuck community. This work is based in part on observations made with the NASA/ESA/CSA James Webb Space Telescope and the NASA/ESA Hubble Space Telescope obtained from the Space Telescope Science Institute, which is operated by the Association of Universities for Research in Astronomy, Inc., under NASA contract NAS 5–26555. The data were obtained from the Mikulski Archive for Space Telescopes at the Space Telescope Science Institute, which is operated by the Association of Universities for Research in Astronomy, Inc., under NASA contract NAS 5-03127 for JWST. These observations are associated with programs \#1727 and \#1837. Some of the data products presented herein were retrieved from the Dawn JWST Archive (DJA). DJA is an initiative of the Cosmic Dawn Center, which is funded by the Danish National Research Foundation under grant No. 140. This work was performed in part at the Aspen Center for Physics, which is supported by the National Science Foundation grant PHY-2210452. This project has received funding from the European Union’s Horizon 2020 research and innovation program under the Marie Sklodowska-Curie grant agreement No 101148925. GEM acknowledges the Villum Fonden research grant 13160 “Gas to stars, stars to dust: tracing star formation across cosmic time,” grant 37440, “The Hidden Cosmos,” and the Cosmic Dawn Center of Excellence funded by the Danish National Research Foundation under the grant No. 140. 
\end{acknowledgments}

\software{\texttt{astropy} \citep{astropy:2013,astropy:2018,astropy:2022}, CASA \citep{McMullin2007ASPC..376..127M}, \texttt{CIGALE} \citep{Boquien2019A&A...622A.103B}, \texttt{Jupyter} \citep{Kluyver2016jupyter}, \texttt{matplotlib} \citep{Hunter:2007}, \texttt{numpy} \citep{harris2020array}}

\appendix

\begin{deluxetable*}{lcr}[ht!]
\tablecaption{\texttt{CIGALE} Parameters \label{tab:cigale}}
\tablehead{
\colhead{Parameter}&
\colhead{Description}&
\colhead{Values}
}
\startdata
& \textbf{Star Formation History (SFH): delayed $+$ burst} & \\
$\tau_{\rm main}$ & e-folding time of main SP model [Myr] & 30, 100, 300, 600, 1000, 3000, 6000 \\
$\tau_{\rm burst}$ & e-folding time of late starburst population model [Myr] & 10000 \\
$f_{\rm burst}$ & mass fraction of late burst population & 0.001, 0.1, 0.3 \\
age & age of main stellar pop [Myr] & 500, 1000, 5000, 10000, 12000 \\
age$_{\rm burst}$ & age of the late burst [Myr] & 50, 100, 300 \\
\hline
& \textbf{Single-age Stellar Population (SSP):} \cite{Bruzual2003MNRAS.344.1000B} & \\
IMF & Initial Mass Function (0 for Salpeter, 1 for Chabrier) & 1 \\
$Z_\odot$ & metallicity & 0.02 \\
\hline
& \textbf{Nebular Emission} & \\
log $U$ & ionization parameter & -3.0 \\
\hline
& \textbf{Dust Attenuation:} \cite{CharlotFall2000ApJ...539..718C} & \\
$A_{V,\rm ISM}$ & V-band attenuation in the ISM & 0, 1, 1.5, 2, 2.5, 3, 4, 5, 6, 7, 8, 9 \\
$\mu$ & ratio of the BC-to-ISM attenuation & 0.1, 0.3, 0.5 \\
\hline
& \textbf{Dust Emission:} \cite{DraineLi2007ApJ...657..810D,Draine2014ApJ...780..172D} & \\
$q_{\rm PAH}$ & mass fraction of PAHs & 0.47, 1.12, 3.9 \\
$u_{\rm min}$ & minimum radiation field & 1, 5, 10, 50\\
\hline
& \textbf{AGN: skirtor2016} & \\
i & inclination angle & 10, 50 \\
fracAGN & AGN fraction & 0.01, 0.1, 0.3 \\
\hline
& \textbf{Radio} & \\
$q_{\rm IR}$ & FIR/radio correlation coefficient & 2.2, 2.6 \\
$\alpha_{\rm SF}$ & slope of power law synchrotron emission from star formation & 0.8 \\
R$_{\rm AGN}$ & radio-loudness parameter & 1, 10, 100 \\
\hline
redshift & & 0.5-8, steps of 0.1
\enddata
\tablecomments{Input parameters adopted by the SCUBADive survey for SED fitting with \texttt{CIGALE}. Parameters not listed adopt the default \cigale\ values.}
\end{deluxetable*}

\bibliographystyle{aasjournal}
\bibliography{scubadiveII}

\end{document}